\documentclass[manuscript]{acmart}

\AtBeginDocument{%
  \providecommand\BibTeX{{%
    \normalfont B\kern-0.5em{\scshape i\kern-0.25em b}\kern-0.8em\TeX}}}
\acmJournal{TQC}
\acmVolume{1}
\acmNumber{1}
\acmArticle{1}
\acmMonth{9}
\setcopyright{none}
\usepackage[utf8]{inputenc}
\usepackage{bbm}

\usepackage{algorithm}
\usepackage[noend]{algpseudocode}
\usepackage{braket}
\newcommand{\vect}[1]{\boldsymbol{\mathbf{#1}}}

\DeclareMathOperator*{\argmin}{arg\,min}

\usepackage{subfig}

\usepackage{xspace}

\usepackage{changes}
\definechangesauthor[name={Ilya}, color=red]{is}
\definechangesauthor[name={Hayato}, color=blue]{hm}
\definechangesauthor[name={Ruslan}, color=orange]{rs}
\definechangesauthor[name={Yuri}, color=purple]{ya}

\usepackage{caption} 
\usepackage{printlen} 
\usepackage[section]{placeins}

\newcommand\compactdots{\makebox[1em][c]{.\hfil.\hfil.}}
\newcommand{\HsubM}{H_{\!M}}
\newcommand{\HsubC}{H_{\!C}}

\makeatletter
\newcommand\thefontsize[1]{{#1 The current font size is: \f@size pt\par}}
\makeatother

\begin{document}
\newpage 
	\title{Multilevel Combinatorial Optimization Across Quantum Architectures}

\author{Hayato Ushijima-Mwesigwa}
\authornote{Both authors contributed equally to this work.}
\authornote{Corresponding authors.}
\authornote{Work on this paper was in part performed while the author was
affiliated with Clemson University. }
\email{hayato@us.fujitsu.com}
\affiliation{%
  \institution{Fujitsu Laboratories of America, Inc.}
  \city{Sunnyvale}
  \state{CA}
  \postcode{94085}
}

\author{Ruslan Shaydulin}
\authornotemark[1]
\authornotemark[2]
\email{rshaydu@g.clemson.edu}
\affiliation{%
  \institution{School of Computing, Clemson University}
  \city{Clemson}
  \state{SC}
  \postcode{29634}
}

\author{Christian F. A. Negre}
\affiliation{%
  \institution{Theoretical Division, Los Alamos National Laboratory}
  \city{Los Alamos}
  \state{NM}
  \postcode{87545}}
\email{cnegre@lanl.gov}

\author{Susan M. Mniszewski}
\affiliation{%
  \institution{Computer, Computational, \& Statistical Sciences Division, Los Alamos National Laboratory}
  \city{Los Alamos}
  \state{NM}
  \postcode{87545}}
\email{smm@lanl.gov}

\author{Yuri Alexeev}
\affiliation{%
 \institution{Computational Science Division, Argonne National Laboratory}
 \city{Argonne}
 \state{IL}
  \postcode{60439}}
  
\author{Ilya Safro}
\authornotemark[2]
\email{isafro@clemson.edu}
\affiliation{%
  \institution{School of Computing, Clemson University}
  \city{Clemson}
  \state{SC}
  \postcode{29634}
}

\renewcommand{\shortauthors}{Ushijima-Mwesigwa and Shaydulin, et al.}

\begin{abstract}
Emerging quantum processors provide an opportunity to explore new approaches for solving traditional problems in the post Moore's law supercomputing era. However, the limited number of qubits makes it infeasible to tackle massive real-world datasets directly in the near future, leading to new challenges in utilizing these quantum processors for practical purposes. Hybrid quantum-classical algorithms that leverage both quantum and classical types of devices are considered as one of the main strategies to apply quantum computing to large-scale problems. In this paper, we advocate the use of multilevel frameworks for combinatorial optimization as a promising general paradigm for designing hybrid quantum-classical algorithms. In order to demonstrate this approach, we apply this method to two well-known combinatorial optimization problems, namely, the Graph Partitioning Problem, and the Community Detection Problem. We develop hybrid multilevel solvers with quantum local search on D-Wave's quantum annealer and IBM's gate-model based quantum processor. We carry out experiments on graphs that are orders of magnitudes larger than the current quantum hardware size, and we observe results comparable to state-of-the-art solvers in terms of quality of the solution.
\\
\noindent {\bf Reproducibility}: Our code and data are available at \cite{code} 
\end{abstract}

\ccsdesc[500]{Mathematics of computing~Graph algorithms}
\ccsdesc[300]{Mathematics of computing~Combinatorial optimization}
\ccsdesc[500]{Hardware~Quantum computation}

\keywords{NISQ, Quantum Annealing, Graph Partitioning, Modularity, Community Detection}

	\maketitle
	\thispagestyle{empty} 

\section{Introduction}

Across different domains, computational 
optimization problems that model large-scale complex systems often introduce a major obstacle to solvers even if tackled with high-performance computing systems. The reasons include a large number of variables and even larger number of interactions, dimensionality required to describe each variable or interaction, and time slices. 
The combinatorial and mixed-integer optimization problems introduce additional layers of complexity, with integer variables often making the problem NP-hard (e.g., in cases of  nonlinearity and nonconvexity). A common practical approach for solving these problems is to use iterative methods.
The iterative methods, while being composed with completely different algorithmic principles, typically share a common property: several fast improvement iterations are followed by a long tail of slow improvement iterations \cite{voss2012meta,kelley1999iterative}. Usually, in such iterative algorithms, solving a large-scale system by using first-order optimization (e.g., gradient descent) or limited observable information (e.g., local search) methods per iteration leads to a local optimum. In other words, local methods tend to converge to a local optimum, which often corresponds to a solution of much lower quality than the true global optimum \cite{horst2000introduction}.  Moreover, in some cases, another problem may exist within each iteration: the algorithms used to solve them are not necessarily exact. In order to accelerate the solvers at each iteration, various heuristics, parallelism-friendly methods, and ad hoc tricks are employed, which often reduce the quality of the solution.



In this paper, we take steps towards building more robust solvers for mid- to large-scale combinatorial optimization problems by fusing two areas whose simultaneous application is only beginning to be explored, namely, quantum computing and multiscale methods. Recent advances in quantum computing provide a new approach for algorithm development for many combinatorial optimization problems. However, Noisy Intermediate Scale Quantum (NISQ) devices are widely expected to be limited to a few hundred, and for certain sparse architectures up to a few thousands qubits. The current state of quantum computing theory and engineering suggests moderately optimistic expectations. In particular, it is believed that in the near future, we will witness relatively robust small-scale architectures with much less noise. This would allow algorithms like the Quantum Approximate Optimization Algorithm (QAOA) and Quantum Annealing (QA) to be run on hardware with minimal error correction efforts. Given the realistic level of precision and, in the case of QAOA, ansatz depth, these algorithms are widely considered to be prime candidates for demonstrating quantum advantage, that is solving a computationally hard problem (such as NP-hard) faster than classical state-of-the-art algorithms. Such algorithms are our first building block.

The multiscale optimization method is our second building block. These methods have been developed to cope with large-scale problems by introducing an approach to avoid entering false local  attraction basins (local optima), a complementary method to stochastic and multistart strategies that help escape it if trapped. Because of historical reasons, on graph problems, they have been termed \emph{multilevel} (rather than multiscale), which we will use here. The multilevel (or multiscale) methods have a long history of breakthrough results in many different optimization problems 
\cite{Cong2003,brandt:review01,migdalas2013multilevel,gelman1990multilevel,KarypisKumar99fast,safro:relaxml,gratton2008recursive,SafroRB08,amg-sss12,leyffer2013fast,hager2018multilevel,Ron2010,shaydulin2019algdist,shaydulin_et_al:LIPIcs:2018:8937,sadrfaridpour2019engineering,sadrfaridpour_ehsan2017} and have been implemented on a variety of hardware architectures. The success of multilevel methods for optimization problems supports our optimism about the ideas proposed in this paper.

No unique prescription exists for how to design multilevel algorithms, but the main idea behind them is to \emph{think globally while acting locally} on a hierarchy of coarse representations of the original large-scale  optimization problem. A multilevel algorithm therefore begins by constructing such a hierarchy of progressively smaller (coarser) representations of the original problem. The goal of the next coarser level in this hierarchy is to approximate the current level problem with a coarser one that has fewer degrees of freedom and thus can be solved more effectively. When the coarse problem is solved, its solution is projected back to the finer level and further refined, a stage that is called uncoarsening. As a result of such a strategy, the multilevel framework is often able to significantly improve the running time and solution quality of optimization methods.  
 The quality of multilevel algorithms in large part depends on that of the optimization solvers applied at all stages of the multilevel framework. In many cases, these locally acting optimization solvers are either heuristics that get stuck in a local optimum or exact solvers applied on a small number of variables (i.e., on subproblems). In both cases, the quality of a global solution can significantly suffer depending on the quality of the solution from the local solver. The optimization algorithms running on the NISQ devices that may replace such local solvers are expected to be a critical missing component to achieve a game-changing breakthrough in multilevel methods for combinatorial optimization. Although the performance of these NISQ-era optimization algorithms is not fully understood (see Sec.~\ref{sec:qopt_scale} for an overview), in this work we do not attempt to rigorously benchmark them. Rather, we focus on the problems arising when integrating these optimization algorithms into a multilevel framework. The proposed synergy between multilevel algorithms and future quantum devices is intended to bridge the gap between slow exact solvers (that are unrealistic to cope with relatively large instances in a reasonable time neither for classical nor NISQ devices) and fast suboptimal heuristics (that are practically employed everywhere) that may never achieve the optimal solution no matter how much time they are given.

In this paper, we introduce Multilevel Quantum Local Search (ML-QLS), which uses an iterative refinement scheme on NISQ devices within a multilevel framework. ML-QLS extends the Quantum Local Search (QLS)~\cite{shaydulin2018community,shaydulin2018network} approach to solve larger problems. This work builds on early results using a multilevel framework and the D-Wave quantum annealer for the graph partitioning \cite{mniszewski2018multilevel}. We demonstrate the general approach  of solving combinatorial optimization problems with NISQ devices in a multilevel framework on two well-known problems. In particular, we solve the Graph Partitioning Problem and the Community Detection Problem on graphs up to approximately $29,000$ nodes using subproblem sizes of 20 and 64 that map onto NISQ devices such as IBM Q Poughkeepsie (20 qubits) and D-Wave 2000Q ($\sim$2048 qubits). Such graphs are orders of magnitude larger than those solved by state-of-the-art hybrid quantum-classical methods. To implement this approach, we develop a novel efficient subproblem formulation method.

In contrast, some of the authors of this paper have previously developed quantum and quantum-classical algorithms for the Graph Partitioning Problem and the Community Detection Problem for multiple parts (> 2) \cite{negre2019detecting,ushijima2017graph}. These did not use a multilevel approach, instead an \emph{all at once} or concurrent approach was employed.

The rest of this paper is organized as follows. In Section \ref{sec:bg}, we discuss the relevant on quantum optimization and multilevel methods, and define the problems. In Sections \ref{sec:methods} and \ref{sec:results}, we discuss the hybrid quantum-classical multilevel algorithm and computational results, respectively. A discussion of the outlook and important open problems that represent major future research directions are presented in Section \ref{sec:conclusion}.

\section{Background}\label{sec:bg}

The methods proposed and implemented in this work aim to solve large graph problems by integrating NISQ optimization algorithms to a multilevel scheme. In this section, we provide a brief introduction into all three components: target graph problems (Sec.~\ref{sec:background:prob_def}), quantum optimization (Sec.~\ref{sec:background:qopt}) and multilevel methods (Sec.~\ref{sec:background:mlcombopt})

Many optimization problems discussed in this work are posed in Ising form. The Ising model is a common mathematical abstraction used to represent the energy of $n$ discrete spin variables $\sigma_i\in \{-1,1\}$, $1\leq i \leq n$,  and interactions $J_{ij}$ between $\sigma_i$ and $\sigma_j$. For each spin variable $\sigma_i$, a local field $h_i$ is specified. The energy of a configuration $\sigma$ is given by the Hamiltonian function: \begin{eqnarray}\label{eq:hamiltonian}
 H(\sigma) = \sum_{i,j} J_{ij}\sigma_i\sigma_j + \sum_i h_i\sigma_i, \quad \sigma_i\in \{-1,1\}.
 \end{eqnarray}
An equivalent mathematical formulation is the Quadratic Unconstrained Binary Optimization (QUBO) problem. The objective of a QUBO problem is to minimize (or maximize) the following function:

\[
H(x) = \sum_{i < j}Q_{ij}x_ix_j + \sum_i Q_{ii} x_i, \quad x\in \{0,1\}.
\]
 
\subsection{Problem Definitions}\label{sec:background:prob_def}

Let $G=(V,E)$ be an undirected graph with vertex set $V$ and edge set $E$. We denote by $n$ and $m$ the numbers of nodes and edges, respectively. For each node $i$, define $\mathbbm{v}_i \in \mathbbm{R}$  as the volume of node $i$ and $A_{ij} \in \mathbbm{R}$ as the positive weight of edge $(i,j)$. For a fixed integer $k$, the \emph{Graph Partitioning Problem} is to find a partition $V_1, \dots, V_k$ of the vertex set $V$ into $k$ parts with equal total node volume such that the total weight of \textit{cut edges} is minimized. A \textit{cut edge} is defined as an edge whose end points are in different partitions. A requirement of equal total sizes  of $V_i$ for all $i$ is sometimes referred as \emph{perfectly balanced} graph partitioning, otherwise an imbalancing parameter is usually introduced to allow imbalanced partitions \cite{bulucc2016recent}. However, in this work we deal with perfect balancing constraints and limit the number of parts to $k=2$. In this case we can write the GP problem as the following quadratic program
\begin{equation}
\begin{aligned}
& \max 
& & \textbf{s}^TA\textbf{s} \\
& \text{s.t.}
& & \displaystyle\sum\limits_{i=1}^n \mathbbm{v}_i s_{i} =0\\
& & & s_{i} \in \{-1,1\},~ i=1 ,\ldots, n,
\end{aligned}
\label{GP_problem}
\end{equation}

\noindent which, as shown in  \cite{ushijima2017graph}, can be reformulated into the following Ising model:
\begin{equation}
    \begin{aligned}
& \max 
& & \textbf{s}^T(\beta A - \alpha \mathbbm{v}\mathbbm{v}^T)\textbf{s} \\
& \text{s.t.}
& &  s_{i} \in \{-1,1\},~ i=1 ,\ldots, n,
    \end{aligned}
    \label{gp_ising}
\end{equation}
for some constants $\alpha, \beta >0$, where $\mathbbm{v}$ is a column vector of volumes such that $(\mathbbm{v})_i = \mathbbm{v}_i$. 

 Maximization of modularity is a famous problem in network science where the goal is to find communities in a network through node clustering (also known as modularity clustering) \cite{Newman06modularity}. For the graph $G$, the problem of Modularity Maximization is to find a partitioning of the vertex set into one or more parts (communities) that maximizes the modularity metric. 
The modularity matrix 
 is a symmetric matrix given by 
  \begin{equation}
    B_{ij}=  A_{ij} - \frac{k_ik_j}{2|E|}, 
    \label{mod_duplicate}
  \end{equation}
  where $k_i$ is the weighted degree of node $i$, namely, $k_i=\sum_j A_{ij}$.  Whereas the modularity is typically defined on unweighted graphs, within the multilevel framework, due to the coarsening of nodes, we primarily work with weighted graphs.
  It can equivalently be written in matrix-vector notation as
  \begin{equation}
      B = A - \frac{1}{2|E|}\mathbbm{k} \mathbbm{k}^T
  \end{equation}
  where $\mathbbm{k}$ is a vector of weighted degrees of the nodes in the graph.  
  For two communities, the \emph{Modularity Maximization Problem}, also referred to as the \emph{Community Detection Problem}, can be written in Ising form as follows:
  
  \begin{equation}
    \begin{aligned}
& \max 
& &  \frac{1}{4|E|}\textbf{s}^T \Big( A - \frac{1}{2|E|}\mathbbm{k} \mathbbm{k}^T \Big )\textbf{s} \\
& \text{s.t.}
& &  s_{i} \in \{-1,1\},~ i=1 ,\ldots, n
    \end{aligned}
    \label{mod_ising}
\end{equation}
where the objective value of (\ref{mod_ising}), for a given assignment of resulting communities, is referred to as the \emph{modularity}.
For more than 2 communities, the Ising formulation of the Community Detection Problem is given in \cite{negre2019detecting}.
  
  Note that the above formulation of Modularity Maximization can be viewed as the Graph Partitioning Problem in the Ising model given in equation (\ref{gp_ising}) where the volume of a node is defined as the weighted degree and the penalty constants $\beta = 1, \alpha = \frac{1}{2|E|}$. We exploit this deep duality between the two problems in our implementation.

\subsection{Optimization on NISQ devices}\label{sec:background:qopt}

In recent years we have seen a number of advances in quantum optimization algorithms that can be run on NISQ devices. The two most prominent ones are the Quantum Approximate Optimization Algorithm (QAOA) and Quantum Annealing (QA), which are inspired by the adiabatic theorem. There are many formulations of the adiabatic theorem (see \cite{albash2a018adiabatic-rs} for a comprehensive review), but all of them stem from the adiabatic approximation formulated by Kato in 1950~\cite{Kato1950}. 
 Adiabatic approximation states, roughly, that a system prepared in an eigenstate (e.g., a ground state) of some time-dependent Hamiltonian $H(t)$ will remain in the corresponding eigenstate\footnote{A note on terminology: a Hamiltonian $H$ is a Hermitian operator. The spectrum of $H$ corresponds to the potential outcomes if one was to measure the energy of the system described by $H$. $\ket{\psi}$ is an eigenstate of a system described by Hamiltonian $H$ with energy $\lambda\in \mathbb{R}$ if $H\ket{\psi} = \lambda \ket{\psi}$. In other words, $\ket{\psi}$ is an eigenvector of $H$ with real eigenvalue $\lambda$.} 
 provided that $H(t)$ is varied ``slowly enough.'' The requirement on the evolution time scales as $O(1/\Delta^2)$ in the worst case~\cite{Elgart2012}, where $\Delta$ is the minimum gap between the ground and first excited state of $H(t)$. 
 
Quantum Annealing is a special case of AQC limited to stochastic Hamiltonians. 
The transverse field Hamiltonian 
\begin{equation}
H_M = \sum_i \sigma^x_i
\end{equation}
is  used as the initial Hamiltonian. The final Hamiltonian is a classical Ising model Hamiltonian with the ground state encoding the solution of the original problem:

\[
H_C = \sum_{ij}J_{ij}s_is_j + \sum_ih_is_i, \quad s_i\in\{-1,+1\}.
\]

The evolution of the system starts in the ground state of $H_M$ and is described by a time-dependent Hamiltonian 
\begin{equation}
    H(t) = \frac{t}{T}H_C + \Big(1-\frac{t}{T}\Big)H_M, ~ t\in (0, T).
\end{equation}

QAOA extends the logic of AQC to gate-model quantum computers and can be interpreted as a discrete approximation of the continuous QA schedule, performed by applying two alternating operators: 
\[W(\beta_k) = e^{-i\beta_kH_M} \text{ and } V(\gamma_k) = e^{-i\gamma_kH_C},
\]
where $W(\beta_k)$ corresponds to evolving the system with Hamiltonian $H_M$ for a period of time $\beta_k$, and $V(\gamma_k)$ corresponds to evolving  $H_C$ for time $\gamma_k$. Similarly to QA, the evolution begins in the ground state of $H_M$, namely, $\ket{+}^{\otimes n}$. Alternating operators are applied to produce the state:


\begin{equation}
   \ket{\psi{(\vect{\beta},\vect{\gamma})}} = e^{-i\beta_p \HsubM}e^{-i\gamma_p \HsubC}\compactdots e^{-i\beta_1 \HsubM}e^{-i\gamma_1 \HsubC}\ket{+}^{\otimes n}= U(\vect{\beta},\vect{\gamma})\ket{+}^{\otimes n}.
\label{eq:ansatz}
\end{equation}

An alternative implementation was proposed, inspired by the success of the Variational Quantum Eigensolver (VQE)~\cite{peruzzo2014variational, Yung2014}. A variational implementation of QAOA combines an ansatz $U(\vect{\beta},\vect{\gamma})$ (that can be different from the alternating operator described above) and a classical optimizer. A commonly used ansatz is a hardware-efficient ansatz~\cite{kandala2017hardware}, consisting of alternating layers of entangling and rotation gates. The algorithm starts by preparing a trial state by applying the parameterized gates to some initial state: $\ket{\psi{(\vect{\beta},\vect{\gamma})}} =  U(\vect{\beta},\vect{\gamma})\ket{+}^{\otimes n}$. In the next step, the state $\ket{\psi{(\vect{\beta},\vect{\gamma})}}$ is measured, and the classical optimization algorithm uses the result of the measurement to choose the next set of parameters $\vect{\beta},\vect{\gamma}$. The goal of the classical optimization is to find the parameters $\vect{\beta},\vect{\gamma}$ corresponding to the optimal QAOA ``schedule,'' that is, the schedule that produces the ground state of the problem Hamiltonian $H_C$:

\begin{equation}
    \vect{\beta_{*}},\vect{\gamma_{*}} = \argmin_{\vect{\beta}, \vect{\gamma}}\bra{\psi{(\vect{\beta},\vect{\gamma})}}H_C\ket{\psi{(\vect{\beta},\vect{\gamma})}}.
\label{eq:argmin_obj_func}
\end{equation}

Both QA and QAOA have been successfully implemented in hardware by a number of companies,  universities, and national laboratories~\cite{dwave2018,iarpaqeo,novikov2018exploring, pichler2018quantum, pagano2019quantum, otterbach2017unsupervised}.

\subsection{On the Scalability of Quantum Optimization Heuristics}\label{sec:qopt_scale}

The question of asymptotic scaling is the central question in the analysis of algorithms. Unfortunately, for many of the most promising quantum optimization algorithms, rigorous analysis (such as provable approximation ratios) beyond the most simple problems is unavailable. Therefore researchers have to resort to experimental evaluations and back-of-the-envelope projections. Such approaches give rise to the second major complication, namely, the fact that empirical results on small problem instances are almost totally uninformative about the overall scaling behavior. 
Famously, the adiabatic quantum algorithm initially appeared to be practically useful for solving NP-complete problems in polynomial time based on numerical simulations of problems of up to tens of variables~\cite{quant-ph/0001106, quant-ph/0104129}. Later analysis, however, has shown that for many problems, both synthetic ones that are classically easy and hard problems like 3-SAT~\cite{vazirani2002gaps}, eigengaps diminish exponentially, leading to exponential worst-case running time for the adiabatic quantum algorithm~\cite{albash2a018adiabatic-rs}. This provides a cautionary example for researchers trying to analyze modern quantum optimization approaches. The exact number of variables needed for the separation between polynomial and exponential scaling to become apparent varies from problem to problem, and the separation might not be clear for small problem instances (i.e., the scaling that is in fact exponential looks polynomial for small instance sizes). However, the increased size of the simulations as well as the hardware (in the case of quantum annealers, reaching thousands of qubits) provides increasing confidence in the potential of quantum optimization heuristics.

Performance on the D-Wave quantum annealer depends on the input, the solver, and the 
hardware platform \cite{mcgeoch2019performance,Denchev2016,King2019,Katzgraber2015}. Changes to the solver include modifying the anneal time or schedule. In this case the platform is the D-Wave 2000Q. Preprocessing strategies such as variable reduction, error mitigation, and improved embeddings applied to the input contribute to more optimized performance. 
The solver tuning strategies include finding an optimal anneal time dependent on the problem, as well as modifying the anneal schedule. Longer anneal times may be required for larger problems. Performance scaling for quantum applications is usually assessed by measuring the dependence of the time to solution (TTS) (for optimized run parameters) on the problem size, as often done for classical applications.

Considering the ground-state success probability for the Sherrington-Kirkpatrick (SK) model and MAX-CUT problems of increasing size was shown to be helpful in understanding the scaling \cite{ Hamerly2019exp}. When evaluating the TTS, one should optimize the run parameters as much as possible, in particular the optimal annealing time. 

Work focused on determining the optimal anneal time on a quantum annealer over classical simulated annealing (SA) for logical-planted instances demonstrated a scaling advantage over SA on the D-Wave quantum annealer \cite{Albash2018-rs}. A scaling advantage in both system size and inverse temperature was demonstrated for simulation of frustrated magnetism in quantum condensed matter for the D-Wave quantum annealer over classical path-integral Monte Carlo \cite{ King2019scale}. An approach to benchmarking quantum annealers using specially crafted problems with varying hardness has been proposed~\cite{Katzgraber2015}. Using a specially crafted problem class, a scaling advantage of algorithms simulating quantum annealing over simulated annealing has been demonstrated numerically for problems too large to be represented on available quantum hardware~\cite{Denchev2016}. This class of problems has been extended and used to show that D-Wave quantum annealer can significantly outperform non-tailored classical state-of-the-art solvers~\cite{King2019}.

A D-Wave API is available for collecting timing information on the details of the Quantum Processing Unit (QPU) 
access time \cite{Dwave2019time}. The QPU access time consists of programming time, sampling time, postprocessing time, and a small amount of time spent for the QPU to regain it's temperature per sample. 
Programming time measures how long it takes to initialize the problem on the QPU. The sampling time is further broken down into per sample times for the anneal, readout, and delay. With this API, runtime scaling for the quantum isomer search problem  
on the D-Wave QA was shown to grow linearly with problem size \cite{Terry2020isomersearch}. In this case the problem size is defined by the number of carbon atoms $n$ in an alkane, which translates to $4(n-2)$ variables; and the number of isomer solutions increases with the number of carbon atoms. 


The scaling of Quantum Approximate Optimization Algorithm (QAOA) is particularly hard due to two factors, namely the presence of outer-loop optimization and the lack of understanding of the scaling of the depth of QAOA circuit required to achieve a good solution. 
The first factor complicating the analysis of QAOA is the outer-loop optimization, i.e., the need to optimize parameters $\vect{\beta},\vect{\gamma}$ in Eq.~\ref{eq:ansatz}. As the landscape is known to be highly non-convex~\cite{Shaydulin2019MultistartDOI, Shaydulin2019EvaluatingDOI,zhou2018quantum}, this optimization becomes a daunting task. Little is known about the structure of this landscape, making it hard to provide an upper bound on the computational hardness of the problem.
At the time of writing the best known upper bound is that the problem of optimizing $\vect{\beta},\vect{\gamma}$ is as hard as finding a global minimizer to a nonconvex problem, where even verifying a feasible point is a local minimizer is NP-hard~\cite{murty1987}. However, in practice a number of techniques have been developed to successfully solve this problem. While the structure is hard to analyze, there have been successful attempts to leverage it using machine learning techniques~\cite{garcia2019quantum,verdon2019learning, khairy2019learning} or accelerate search with multi-start optimization \cite{Shaydulin2019MultistartDOI}. These results make us hopeful that with the help of a pre-trained model, high-quality QAOA parameters can be found in a small number of objective evaluations. There have been promising results showing that in higher-depth regime for some problems it is possible to avoid optimization altogether and use a smoothly extrapolated set of parameters reminiscent of an adiabatic schedule~\cite{zhou2018quantum}.

The second factor in the lack of analytical and empirical results on QAOA behavior in low- to medium-depth regime (e.g., $5 \leq p\leq 100$). Analytically, QAOA appears to be hard to analyze beyond $p=\{1,2\}$ for non-trivial problems~\cite{wang2018quantum, brandao2018fixed,Szegedy2019qaoaenergies}. At the same time, even for very simple problems and small instance sizes it is clear that achieving a good solution requires going beyond at least $p=5$~\cite{Shaydulin2019EvaluatingDOI, crooks2018performance, Szegedy2019qaoaenergies}. Therefore we have to rely on empirical results to answer the question of exactly how large the $p$ needs to be in order to achieve a good solution. This empirical evaluation is impeded by the complexity of simulating QAOA in medium depths. On one end, traditional state-vector based simulators have running time exponential in number of \emph{qubits}, limiting the problem sizes to tens of variables. On the other end, tensor network based simulators have running time that is exponential in the number of \emph{gates}, limiting the depth of the QAOA circuit that can be simulated. At the time of the writing, the state-of-the-art simulators were limited to simulating a thousand qubits to depth $p=5$~\cite{huang2019alibabacloud}. These two constraints (on number of qubits for one simulation approach and on the depth for the other) make it challenging to numerically analyze QAOA performance in the crucial zone of medium-sized problems (hundreds to thousands of variables) and medium-depth circuit ($p>10$). The high levels of noise and small number of available qubits make this analysis impossible on the currently available quantum hardware. All of the aforementioned complications contribute to the lack of the results showing how QAOA depth $p$ scales with the size of the problem. At the same time, results presented in~\cite{crooks2018performance, zhou2018quantum} indicate that at least for the problem sizes small enough to fit on near-term quantum computers, $10\leq p\leq 30$ is sufficient to achieve high-quality solutions.

Due to the limitations of the available hardware, in this paper we do not run full QAOA ansatz on the IBM Q hardware. Instead, we use a shallow-depth hardware-efficient ansatz (HEA)~\cite{kandala2017hardware}. Much less is known about the potential of such ansatzes to produce quantum speedup. Numerical experiments on small problem instances do not show that quantum entanglement provides an advantage in optimization~\cite{Nannicini2019performance}. At the same time, recent analytical results show that HEAs are efficiently simulatable classically in constant depth~\cite{napp2019efficientshallow} and suffer from exponentially vanishing gradients in polynomial depth~\cite{McClean2018barrendoi} (here depth is a function of number of qubits). A recent result shows that in logarithmic depth regime gradient vanishes only polynomially, making HEAs trainable in this regime~\cite{cerezo2020costfunctiondependent}. This result indicates a potential for quantum advantage using HEAs in logarithmic depth, as they are both hard to simulate classically and do not suffer from exponentially vanishing gradients. Evaluating the potential for quantum advantage from using logarithmic-depth HEAs in QAOA setting for optimization remains an open problem.
  
\subsection{Multilevel Combinatorial Optimization}\label{sec:background:mlcombopt}

\begin{figure*}[b]
    \centering
    \includegraphics[width=\textwidth, trim=0cm 13cm 0cm 0cm]{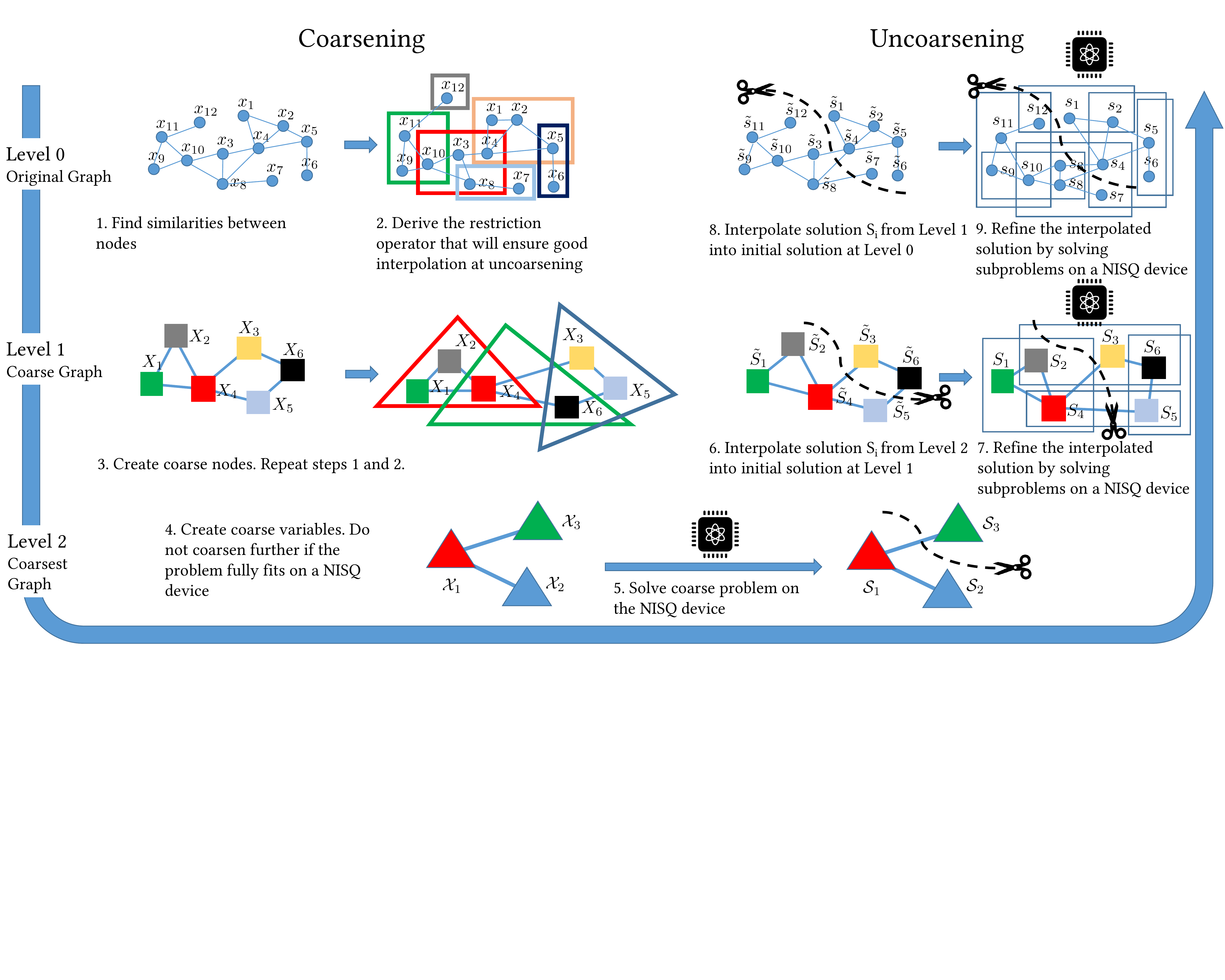}
    \caption{V-cycle for a graph problem. First, the problem is iteratively coarsened (left). Second, the coarse problem is solved using a NISQ optimization solver (bottom). Finally, the problem is iteratively uncoarsened and the solution is refined using a NISQ solver (right).}
    \label{fig:multilevel}
\end{figure*}

The goal of the multilevel approach for optimization problems on graphs is to create a hierarchy of coarsened graphs $G_0=G$, $G_1$, ... ,$G_k$ in such a way that the next coarser graph $G_{i+1}$ ``approximates''  some properties of $G_i$ (that are directly relevant to the optimization problem of interest)  with fewer degrees of freedom. After constructing such a hierarchy, the coarsening is followed by solving the problem on $G_k$ as best as we can (preferably exactly) and then the uncoarsening projects the solution back to $G_0$ through gradual refinement at all levels of the hierarchy, with a refined solution at level $i+1$ serving as the initial solution at level $i$. The entire coarsening-uncoarsening process is called a V-cycle. Other variations of hierarchical coarsening-uncoarsening schemes include W- and F-cycles \cite{vlsicad}. Figure~\ref{fig:multilevel} presents an outline of a V-cycle.

Typically, when solving problems on graphs in which  nodes represent the optimization variables (such as those in the partitioning and community detection),  having fewer degrees of freedom implies a decreased number of nodes in each next coarser graph:
$|V_0| > |V_1| > |V_2| > ... > |V_k|$.\footnote{Note that this does not necessarily implies $|E_0| > |E_1| > |E_2| > ... > |E_k|$}
With a smaller number of variables at progressively coarser levels, one can use more sophisticated but possibly slower refinement algorithms. 
However, it is still not sufficient to solve the whole  problem until the coarsening reaches the coarsest level.
As a result, at each level, the actual solution is produced by a refinement. Refinement is typically implemented with a decomposition method that uses a previous iteration or a coarser-level solution as an initialization. The multilevel algorithms rely heavily \cite{Walshaw2004} on the quality of refinement solvers for small and local subproblems at all levels of coarseness. 
Thus, the most straightforward way to use NISQ devices in multilevel frameworks is to iteratively apply them as local solvers to refine a solution inherited from the coarse level. Because the refinement is executed at all levels of coarseness, \emph{even a small improvement of a solution at the coarse level  may cause a major improvement at the finest scale}. 
Typically, this is the most time-consuming stage of the multilevel solvers  which is expected to be fundamentally better if improved by NISQ devices. 

Most refinement solvers in multilevel frameworks rely on fast  but low-quality heuristics, rather than on the ability to compute an optimal solution. Moreover, in many existing solvers, the number of variables in such local subproblems is comparable with or smaller than the the size of the problems that can be directly embedded on the NISQ devices (see examples in \cite{hager2018multilevel,leyffer2013fast}), making them a perfect target for NISQ optimization algorithms.
In most multilevel/multiscale/multigrid-based optimization solvers, a refinement consists of covering the domain (or all variables) with \emph{small} subsets of variables (i.e., small subproblems) such that solving a small local problem on a subset improves the global solution  for the current level.


Multilevel graph partitioning and community detection algorithms are examples of the most successful applications of multilevel algorithms for large graphs, achieving an excellent time/quality trade-off \cite{bulucc2016recent}. 
In this paper, we deliberately use the simplest version of coarsening (in order to focus on the hybrid quantum-classical refinement) in which the edges of the fine level graph are collapsed 
 and create coarse level vertices by merging the fine level ones. 
 There are several classes of refinement for both problems but in all of them, at each step a small subset of nodes (or even a singleton) is reassigned with partition (or cluster) that either better optimizes the objective or improves constraints. Some variants of stochastic extensions also exist.
 

  
\subsection{On Scalability of Multilevel Methods}\label{sec:multilevel_scale}

If we do not consider algorithmic frameworks with very limited space (such as streaming), the scalability of a correctly designed in-memory multilevel framework with its instance graphs is practically limited (or proportional) to the available memory size. Requirements to keep graphs in memory (not necessarily RAM) for multilevel partitioning and community detection are similar to those of matrices for multigrid \cite{vlsicad}, so the complexity is also comparable up to the factor of refinement. Theoretically, the multilevel and multigrid frameworks exhibit $O(|E|)$ or $O(\text{number of non-zeros in a matrix})$ complexity. However, for optimization on graphs, the refinement stage is typically computationally more expensive than that for the multigrid (e.g., compare Gauss-Seidel relaxation sweeps \cite{livne2012lean} and Kernighan-Lin or min-cut/max-flow refinement in graph partitioning \cite{bulucc2016recent}) because at each step of the refinement, an integer problem has to be solved. Refinement for the relaxed versions of integer problems (e.g., for the minimum 2-sum or bandwidth \cite{SafroRB08}) are usually faster but the quality suffers and in the end they should follow some rounding scheme for integer solution. However, even if the the refinement complexity is linear in the number of edges or corresponding matrix non-zeros, some overhead is typically introduced for the integer problems.

In this paper, we deliberately use a simple coarsening which folds edges by merging pairs of nodes. The situation with the scalability of multilevel frameworks if high-order interpolation coarsening is involved (e.g., algebraic multigrid inspired weighted aggregation \cite{Safro2006} when nodes can be split into several fractions, and different fractions form coarse nodes) is different. The high-order interpolation coarsening may result in increasing number of edges at several fine levels immediately implying increased running time. In such cases, the complexity can increase to $\max_{\text{level }i}O(E_i)$. Subsequently, a larger graph at each level requires more intensive refinement. In addition, the number of refinement calls required to achieve a very good solution strongly depends on the coarsening quality which makes it difficult to get a complexity required for a nearly optimal solution. The only practical solution to that is artificially limiting the number of refinement calls (see all major solvers such as Metis, Jostle, Kahip, and Scotch reviewed in \cite{bulucc2016recent}). 

The criteria for limiting the number of refinement calls is never ideal. The refinemnt algorithms always heuristically decide what vertex or group of vertices should be optimized with respect to the current assignment of vertices to clusters (or parts) to make the current solution better. Typically, they take so called ``boundary'' nodes in all parts, i.e., those whose move from part to part can potentially improve the solution and optimize them or their groups. Therefore, the scalability of multilevel graph partitioning and clustering frameworks can be loosely described in terms of $|V|$ and $|E|$. Instead, it is better to describe it in terms of expected cut similar to the analysis in \cite{KarypisKumar95b}. To the best of our knowledge, there is no analysis that connects the size of the graph and performance of multilevel algorithms providing any practically useful theoretical bounds. Currently, the best way to describe the scalability of multilevel graph partitioning and clustering solvers is to use $O(|E|)$. However, the hidden constant in the linear complexity depends on the type of refinement. Thus, we anticipate that with the advent of reliable quantum hardware, one can expect a significant improvement in the running time and quality of the refinement in multilevel frameworks which will eliminate computationally expensive solvers to locally optimize groups of variables.  
We refer the reader to one of the most recent examples of multilevel scalability in \cite{davis2019algorithm} (e.g., see Table IV) in which a graph with 3.8B edges was (suboptimally) solved in 255 seconds.
    
\section{Methods}\label{sec:methods}

An iterative improvement scheme is a common approach for solving large scale problems with NISQ devices. Traditionally, this is done by formulating the entire problem in the Ising model or as a QUBO and then solving it using hybrid quantum-classical algorithms (see, for example, "qbsolv" from D-Wave systems \cite{booth2017partitioning}). These methods decompose the large QUBO into smaller sub-QUBOs or decrease the number of degrees of freedom to fit the subproblem on the hardware (for example, using a multilevel scheme), and iteratively improve the global solution by solving the small subproblems (sub-QUBOs). One of the main limitations of this approach is the size and density of the original QUBO. For example, in the graph partitioning formulation given by Equation~\ref{gp_ising}, the term $\mathbbm{v}\mathbbm{v}^T$ leads to the formulation of a completely dense $n \times n$ QUBO matrix regardless of whether or not the original graph was sparse. Storing and processing this dense matrix can easily make this method prohibitively computationally expensive even for moderately sized problems. In our implementation of Quantum Local Search (QLS)~\cite{shaydulin2018network} we circumvent this limitation by developing a novel subproblem formulation of the Graph Partitioning Problem and Modularity Maximization as a QUBO that does not require formulating the entire QUBO.

Another concern is the effectiveness of selection criteria of candidate variables (or nodes) to be included in each subproblem. A common metric used in selecting whether or not a variable is to be included in the subproblem is whether or not changing the variable value would reduce (increase) the objective value for a minimization (maximization) problem. Thus, since computing the change in objective value for a small change in the solution is performed multiple times, it is important to ensure that this computation is efficient. We derive a novel efficient way to compute the change in the objective value of the entire QUBO also without formulating the entire QUBO and thus provide an efficient refinement scheme using current NISQ devices. 

We begin by introducing an efficient QUBO subproblem formulation for the Graph Partitioning Problem, and the Community Detection Problem. Then we present an efficient way to compute the gain and change in the objective of the entire QUBO. Finally, we put it all together and outline our algorithm.

\subsection{QUBO formulation for subproblems}
Let $M$ be an $n \times n$ symmetric matrix that represents the QUBO for a large scale problem such that it is prohibitively expensive to either generate or store $M$. However, for QLS we need to generate constant-size sub-QUBOs of $M$ which in turn represent subproblems of the original problem. In order to generate a sub-QUBO, let $k$ be the size of the desired sub-QUBO. In other words, the sub-QUBO will have $k$ variables and $n-k$ \emph{fixed variables} that remain invariant for this specific sub-QUBO. We refer to the $k$ variables as \emph{free variables}.  Without loss of generality, let the the first $k$ variables of $\textbf{s}$ be the free variables, then we write $\textbf{s}$ as 

  \begin{equation*}
    \textbf{s} = \begin{bmatrix}
    \textbf{s}_v  \\
   \textbf{s}_f \\
    \end{bmatrix},
  \end{equation*}
where $\textbf{s}_v$ represents the $k$ free variable terms and $\textbf{s}_f$ represents the $n-k$ fixed terms.  In the next step, $M$ can be represented using block form
\begin{equation}
   M=\left[
   \begin{array}{c|cccc}
~ \large M_{vv}  ~&& &\large M_{vf} \quad &~\\\hline
&&&&~\\
&&&&~\\
\large M_{vf}^T& && \large M_{ff}\\
&&&&~\\
&&&&~\\
\end{array}
\right] 
\label{block}
\end{equation}
such that $ M_{vv} $  is a $k \times k$ matrix. Next, we can write $\textbf{s}^T M \textbf{s}$ as 
\begin{equation}
 \textbf{s}^T M\textbf{s} =  \textbf{s}_v^T M_{vv}  \textbf{s}_v + \textbf{s}_v^T (2M_{vf}  \textbf{s}_f) + \textbf{s}_f^T M_{ff}  \textbf{s}_f
\end{equation}
Since $\textbf{s}_f$ are fixed values, we have $\textbf{s}_f^T M_{ff}  \textbf{s}_f$  
as a constant thus

\begin{equation}
    \min  \ \textbf{s}^T M\textbf{s} =  \min \   \textbf{s}_v^T M_{vv}  \textbf{s}_v + \textbf{s}_v^T (2M_{vf}  \textbf{s}_f) 
    \label{variable}
\end{equation}
From equation (\ref{block}), we have
\begin{equation}
  \mathbbm{v}\mathbbm{v}^T=\left[
   \begin{array}{c|cccc}
~ \mathbbm{v}_v\mathbbm{v}_v^T  ~&& &\mathbbm{v}_v\mathbbm{v}_f^T \quad &~\\\hline
&&&&~\\
&&&&~\\
\mathbbm{v}_f\mathbbm{v}_v^T& && \mathbbm{v}_f\mathbbm{v}_f^T\\
&&&&~\\
&&&&~\\
\end{array}
\right] 
\end{equation}
Therefore, from equation (\ref{variable}), we have
\begin{equation}
\min  \ \textbf{s}^T  \mathbbm{v}\mathbbm{v}^T\textbf{s}  = \min  \ \textbf{s}_v^T \mathbbm{v}_v \mathbbm{v}_v^T\textbf{s}_v + 2\textbf{s}_v^T \mathbbm{v}_v\mathbbm{v}_f^T   \textbf{s}_f 
\label{sub_vol}
\end{equation}
The formulation in (\ref{sub_vol}) is particularly important because it shows that the matrix $\mathbbm{v}\mathbbm{v}^T$ does not need to be explicitly created at each iteration during refinement. This is a crucial observation because $\mathbbm{v}\mathbbm{v}^T$ is a completely dense matrix. 

  
  As described in Sec.~\ref{sec:background:prob_def}, the Community Detection Problem is given by
    \begin{equation}
     \max \ \frac{1}{4|E|}\textbf{s}^T \Big( A - \frac{1}{2|E|}\mathbbm{k} \mathbbm{k}^T \Big )\textbf{s}
  \end{equation}
  or 
      \begin{equation}
     \min \ \textbf{s}^T \Big(\frac{1}{2|E|}\mathbbm{k} \mathbbm{k}^T - A \Big )\textbf{s}
  \end{equation}
  and the Graph Partitioning Problem is given by
       \begin{equation}
     \min \ \textbf{s}^T \Big(\alpha \mathbbm{v} \mathbbm{v}^T - \beta A \Big )\textbf{s}.
  \end{equation}
  In the above formulation, modularity clustering can be viewed as the Graph Partitioning Problem in a QUBO model, where the volume of a node is defined as the weighted degree and the penalty constant is $\frac{1}{|E|}$
  Therefore, in both cases we can perform a refinement while  defining fixed values as
  \begin{equation}
          \begin{aligned}
     \min \ \textbf{s}^T \Big(\frac{1}{2|E|}\mathbbm{k} \mathbbm{k}^T - A \Big )\textbf{s} &= \min \  \textbf{s}_v^T \Big (\frac{1}{2|E|} \mathbbm{k}_v \mathbbm{k}_v^T \Big )\textbf{s}_v +  \textbf{s}_v^T \Big (\frac{1}{|E|} \mathbbm{k}_v\mathbbm{k}_f^T \Big )  \textbf{s}_f  - \textbf{s}^TA \textbf{s}  
  \end{aligned}
  \label{eq:mod_refine}
  \end{equation}

  and
  \begin{equation}
\begin{aligned}
     \min \ \textbf{s}^T \Big(\alpha \mathbbm{v} \mathbbm{v}^T - \beta A \Big )\textbf{s} &= \min \  \textbf{s}_v^T  \Big ( \alpha \mathbbm{v}_v \mathbbm{v}_v^T \Big )\textbf{s}_v  +  \textbf{s}_v^T \Big  (2\alpha \mathbbm{v}_v\mathbbm{v}_f^T  \Big ) \textbf{s}_f  - \beta \textbf{s}^TA\textbf{s} 
  \end{aligned}
       \label{eq:gp_refine}
  \end{equation}

  with
  \begin{equation}
      \min \ - \beta\textbf{s}^TA \textbf{s} = \min \ - \beta\textbf{s}_v^TA_{vv} \textbf{s}_v - \textbf{s}_v^T (2\beta A_{vf}\textbf{s}_f)
  \end{equation}
  The formulation in (\ref{eq:mod_refine}) and (\ref{eq:gp_refine}) are particularly important during the refinement step because this implies that the complete dense (and therefore prohibitively large) QUBO or Ising model does not need to be created at each iteration. These formulations also demonstrate a close relationship between the Graph Partitioning Problem and the Community Detection Problem.

\subsection{Efficient Evaluation of the Objective}
  In order to select the free variables for the subproblem, we need to be able to efficiently compute the change of the objective function by moving one node from one part to another. 
In other words, for each vertex $v$, we need to efficiently compute the  \emph{gain}, which is the decrease (or
increase) in the edge-cut together with penalty if $v$ is moved to the other part. 

For a symmetric matrix $M$, the change in the value  $Q = \textbf{s}^TM \textbf{s}$ by flipping a single variable $s_i$ corresponding to the node $i$ is given by
\begin{equation}
    \Delta Q(i) = 2(\sum_{j \in C_1} M_{ij} - \sum_{j \in C_2}M_{ij})
\end{equation}
where $C_1$ and $C_2$ correspond to all variables with $s_i = -1$ and $s_i = 1$ respectively. 
Next, we define
\begin{equation*}
    \begin{aligned}
         deg(v, C) := \sum_{j \in C} A_{vj}; & \text{~}&
         Deg(C) := \sum_{i \in C} k_i;  & \text{~}&
         Vol(C) := \sum_{i \in C} \mathbbm{v}_i
    \end{aligned}
\end{equation*}
then
\begin{align*}
    2(\sum_{j \in C_1} A_{ij} - \sum_{j \in C_2}A_{ij}) &= 2deg(v_i, C_1) - 2deg(v_i, C_2)
\end{align*}
and finally
\begin{align*}
    2(\sum_{j \in C_1} (\mathbbm{v} \mathbbm{v}^T)_{ij}  -  \sum_{j \in C_2}(\mathbbm{v} \mathbbm{v}^T)_{ij}) &= 2\Big (\mathbbm{v}_i \sum_{j \in C_1, i \neq j} \mathbbm{v}_j - \mathbbm{v}_i \sum_{j \in C_2}\mathbbm{v}_j \Big )\\
    & = 2 \mathbbm{v}_i \big (Vol(C_1\backslash i)   - Vol(C_2) \big ),
\end{align*}
where we assume that $i \in C_1$. This expression can be computed in $O(1)$ time. 

In the same way

\begin{align*}
    2(\sum_{j \in C_1} (\mathbbm{k} \mathbbm{k}^T)_{ij} - \sum_{j \in C_2}(\mathbbm{k} \mathbbm{k}^T)_{ij})&= 2\Big (k_i \sum_{j \in C_1, i \neq j} k_j - k_i \sum_{j \in C_2}k_j \Big )\\ 
    & = 2 k_i \big (Deg(C_1\backslash i) - Deg(C_2) \big )
\end{align*}
can also be computed in $O(1)$ time given $ Deg(C_1) $ and $ Deg(C_2) $, where $ Deg(C_i) $ represents the sum of weighted degrees of nodes in community $i$.

Therefore, the change in modularity is given by 
\begin{equation}
    \begin{aligned}
    \Delta Q(i) &= \frac{ k_i}{|E|} \big (Deg(C_1\backslash i) - Deg(C_2) \big )- 2\Big (  deg(v_i, C_1) - deg(v_i, C_2) \Big )
\end{aligned}
\label{mod_update}
\end{equation}

and change in edge-cut together with penalty value is given by 
\begin{equation}
    \begin{aligned}
    \Delta Q(i) &= 2 \alpha  \mathbbm{v}_i \big (Vol(C_1\backslash i)  - Vol(C_2) \big ))- 2\beta \Big (  deg(v_i, C_1) - deg(v_i, C_2) \Big ) 
\end{aligned}
     \label{gp_update}
\end{equation}

For each node $i$, both expressions (\ref{mod_update}) and (\ref{gp_update}) can be computed in $O(k_i)$ time, where $k_i$ is the unweighted degree of $i$.

At no point during the algorithm should the complete QUBO matrix be formulated. This also applies to the process of evaluating a given solution. In other words, evaluating the modularity for the Community Detection Problem or edge-cut together with penalty term for the Graph Partitioning Problem should be done in $O(1)$ time and space. 
The term is
\begin{equation*}
      \textbf{s}^T \mathbbm{v}\mathbbm{v}^T\textbf{s} = \big (Vol(C_1) - Vol(C_2) \big )^2
\end{equation*}
where as
\begin{equation*}
      \textbf{s}^T A\textbf{s} = 2(|E| - 2cut).
\end{equation*}
Therefore, 
\begin{equation}
\begin{aligned}
     \textbf{s}^T( \alpha\mathbbm{v}\mathbbm{v}^T- \beta   A)\textbf{s} &= \alpha \Big (Vol(C_1) - Vol(C_2) \Big )^2 - 2\beta(|E| - 2cut)
\end{aligned}
\label{no_qubo_gp_en}
\end{equation}
and
\begin{equation}
\begin{aligned}
         \textbf{s}^T \Big(  \frac{1}{2|E|}\mathbbm{k} \mathbbm{k}^T  - A\Big )\textbf{s} &= \frac{1}{2|E|}\Big (Deg(C_1) - Deg(C_2) \Big )^2 - 2(|E| - 2cut)
\end{aligned}
\label{no_qubo_mod_en}
\end{equation}
where equations (\ref{no_qubo_gp_en}) and (\ref{no_qubo_mod_en}) give the formulations for computing the modularity and edge-cut with corresponding penalty value respectively without creating the QUBO matrix. 

\subsection{Algorithm Overview}

Now we can combine the building blocks described in the previous two subsections. Let $G=(V,E)$ be the problem graph. ML-QLS begins by coarsening the problem graph. 
During the coarsening stage, for some integer $k$, a hierarchy of coarsened graphs $G = G_0, G_1, \dots ,G_k$ is constructed. In this work, we used the coarsening tools implemented in KaHIP Graph Partitioning package~\cite{sandersschulz2013}. We used the coarsening implementation that is performed using maximum weight matching with ``expansion$^{*2}$'' metric as described in~\cite{Holtgrewe2010}. The maximum edge matching is found using the Global Path Algorithm ~\cite{Holtgrewe2010}. In the next step, a QUBO is formulated for the smallest graph $G_k$ and solved on the quantum device. If $|V_k|$ is greater than the hardware size\footnote{more specifically, greater than the maximum number of variables in a problem that can be embedded on the device}, QLS~\cite{shaydulin2018network} with a random initialization is used to solve for $G_k$. Then, the solution is iteratively projected onto finer levels and refined using QLS. The algorithm overview is presented in Alg.~\ref{alg:outline}.

For the Graph Partitioning Problem, the initial weight of each node is one by definition, therefore coarsening of the nodes keeps the total node volume constant at each coarsening level.  For the Community Detection Problem, the initial weight of each node is set to the degree of the node. This ensures that the size of the graph (total number of weighted edges) is also kept constant at each level. Note that Graph Partitioning is defined with respect to total node volume ($|V|)$, while modularity is defined with respect to the size ($|E|$, the total number of weighted edges) of the graph.

\begin{algorithm}[tbp]
 \caption{Multilevel Quantum Local Search}\label{alg:outline}
\begin{algorithmic}
\Function{ML-QLS}{$G$, problem\_type}
\If{problem\_type is modularity}
\State $G$ = UpdateWeights($G$)
\EndIf 
\State $G_0, G_1, \ldots ,G_k$ = KaHIPCoarsen($G$)
\If{$|V_k|\leq $HardwareSize}
\State // \textit{solve directly}
\State QUBO = FormulateQUBO($G_k$)
\State solution = SolveSubproblem(QUBO)
\Else
\State // \textit{use QLS}
\State initial\_solution = RandomSolution($G_k$)
\State solution = RefineSolution($G_k$, initial\_solution)
\EndIf
\For{$G_i$ in $G_{k-1}, G_{k-2}, \ldots ,G_0$}
\State projected\_solution = ProjectSolution(solution, $G_i$, $G_{i+1}$)
\State solution = RefineSolution($G_i$, projected\_solution)
\EndFor
\Return solution
\EndFunction
\end{algorithmic}

\begin{algorithmic}
\Function{RefineSolution}{$G_i$, projected\_solution}
\State solution = projected\_solution
 \While{not converged}
    \State $\Delta Q$ = ComputeGains($G_i$, solution)
  	\State $X$ = HighestGainNodes($\Delta Q$)
  	\State QUBO = FormulateQUBO($X$)
  	\State // \textit{using IBM UQC or D-Wave QA}
  	\State candidate = SolveSubproblem(QUBO)
  	\If{$\mbox{candidate} > \mbox{solution}$}
  	\State solution = candidate
    \EndIf
 \EndWhile
 
 \Return solution
\EndFunction
\end{algorithmic}
\end{algorithm}

\begin{figure*}[ht]
    \subfloat[Edge weights of the coarsest graph \texttt{SSS12}.]{\includegraphics[width=0.5\textwidth]{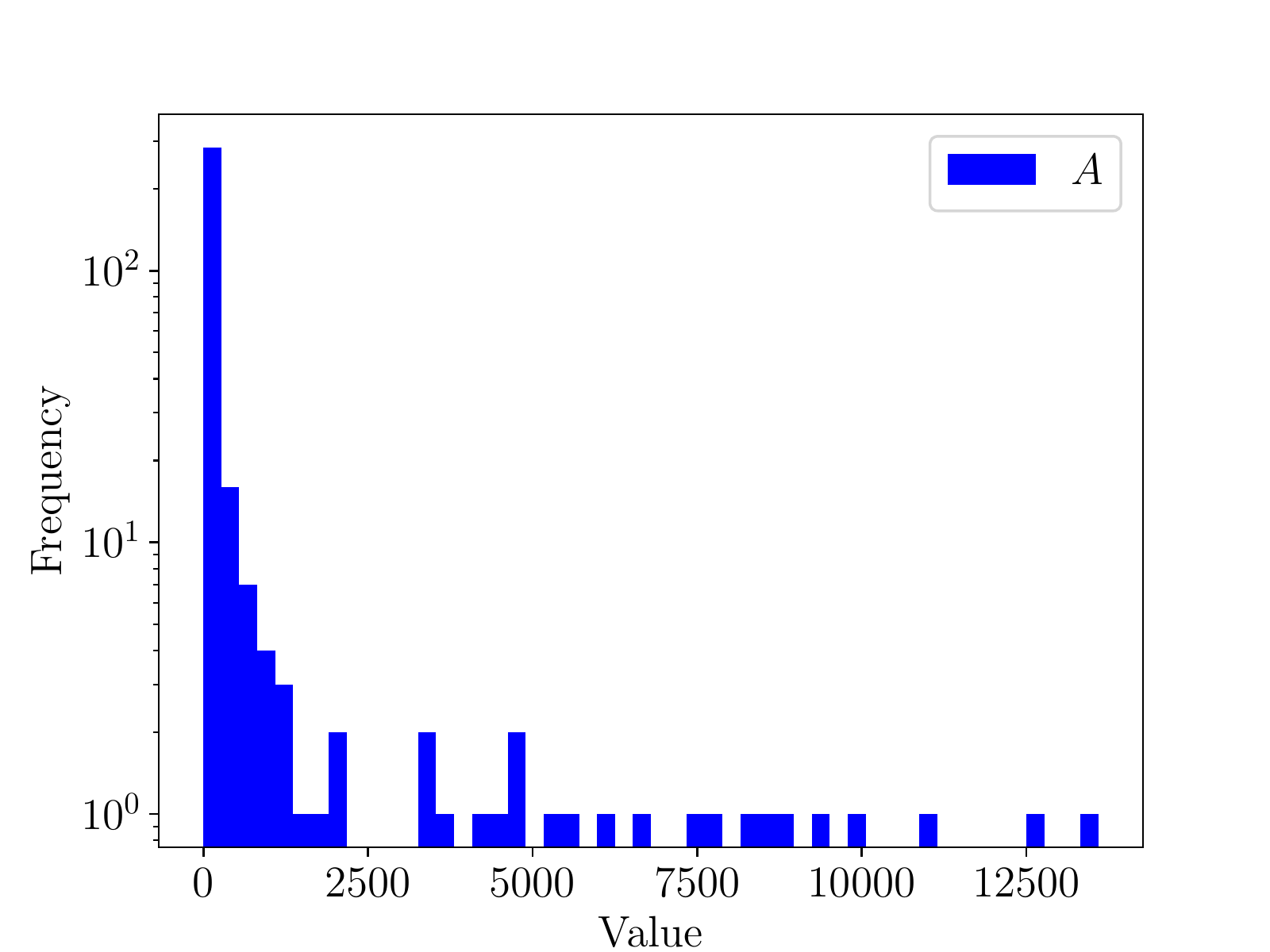}%
    \label{fig:edgeweird_sss12}}
    \hfill
    \subfloat[Entries of the matrix $\mathbbm{v}\mathbbm{v}^T$.]{\includegraphics[width=0.5\textwidth]{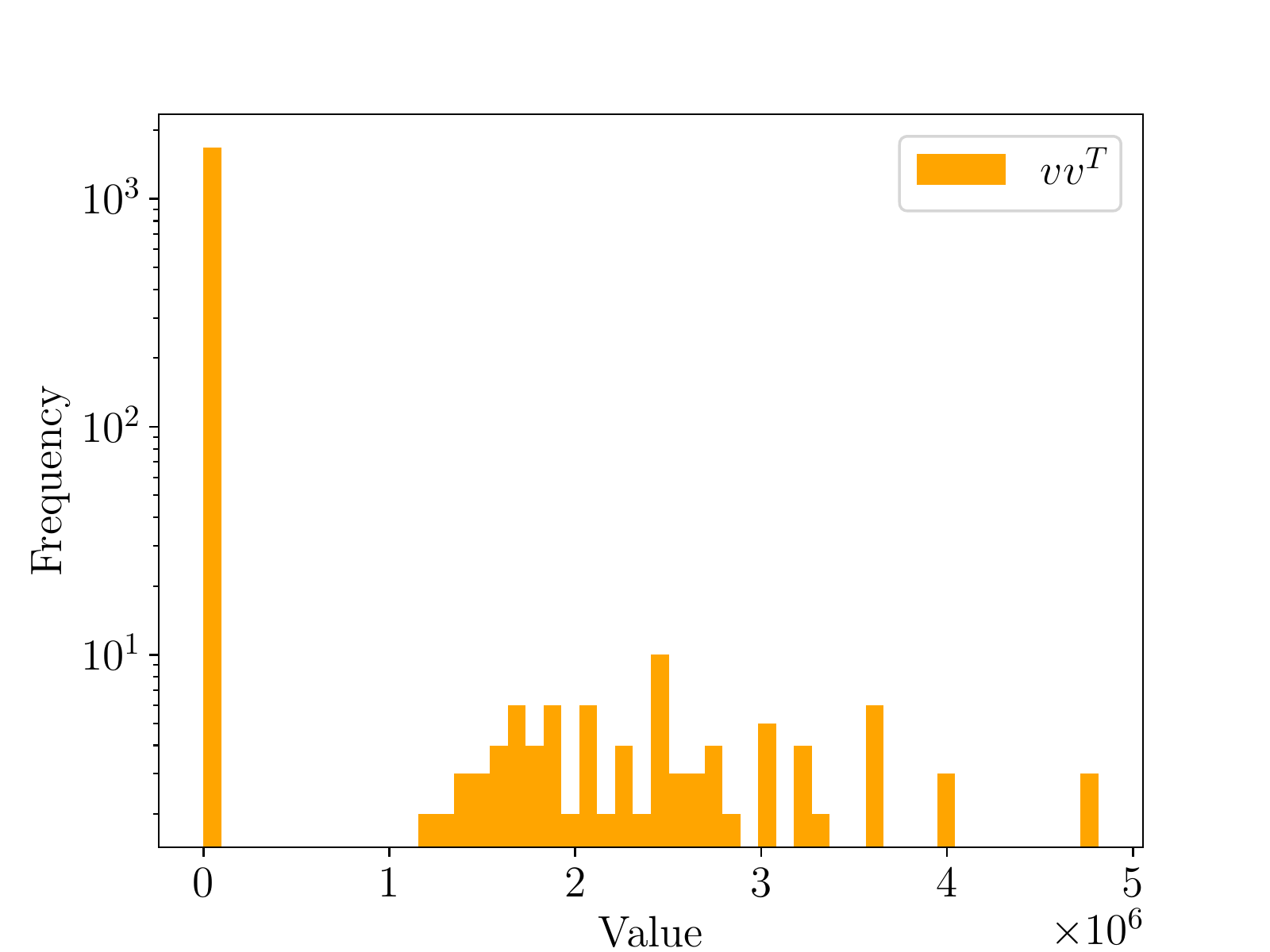}%
    \label{fig:vvT}}
    \caption{In Figure~\ref{fig:edgeweird_sss12}, the maximum value is approximately $13 \times 10^{3}$.In Figure~\ref{fig:vvT}, the maximum value is approximately $5 \times 10^6$ and minimum value $1$. 
    A naive scaling of QUBO matrix $A - \mathbbm{v}\mathbbm{v}^T$ can result in values that are too large to be handled by the quantum annealer due to its limited precision. Such values of $A$ are ignored, leading to random balanced partitions.}
\end{figure*}

\subsection{Addressing the Limited Precision of the Hardware}

One of the subproblem solvers we used in this work is Quantum Annealing, which we ran on the LANL D-Wave 2000Q machine. The D-Wave 2000Q is an analog quantum annealer with limited precision. 
In this work, we used a simple coarsening that constructs coarser graphs by aggregating nodes at a finer level to become a single node at the coarser level (i.e. many nodes on the finer level are merged into one node at the coarser level, with the volume of the new node set to be the sum of the volumes of the nodes on the coarser level). This causes the precision required to describe the node volumes and edge weights for coarser graphs to increase dramatically, especially for the large scale problems.
Thus, a QUBO describing the coarsest graph could require significantly more precision to represent compared to the finest graph. For example, in Graph Partitioning where the QUBO problem to be minimized is $A - \alpha \mathbbm{v}\mathbbm{v}^T$, the range of values in the matrix $A$ increase at a different rate than the range of values in the matrix $\mathbbm{v}\mathbbm{v}^T$ during the coarsening process, increasing the precision required to describe the overall QUBO formed at each level (see an example on Fig.~\ref{fig:edgeweird_sss12}). Thus, if the QUBO $A - \alpha \mathbbm{v}\mathbbm{v}^T$ is directly scaled to accommodate the limited precision of the device, the quality of the results can suffer. In our experiments, we observe that directly scaling the QUBO returned feasible, but low quality solutions.  In order to overcome this challenge, for the problems solved on the D-Wave device, we first scaled the matrices $A$ and $\alpha \mathbbm{v}\mathbbm{v}^T$ separately, and then formed the QUBO to be optimized. This approach then resulted in achieving results with high quality solutions on the D-Wave device. 

\section{Experiments and Results}\label{sec:results}
\paragraph{Implementation}The general framework for ML-QLS is implemented in Python 3.7 with NetworkX~\cite{hagberg2008-rs} 
for network operations. We have used the coarsening algorithms available in the KaHIP Graph Partitioning package~\cite{sandersschulz2013} which are implemented in C++. The code for the general ML-QLS framework is available on GitHub~\cite{code}.

\paragraph{Systems} The refinement algorithms presented in this work require access to NISQ devices capable of solving problems formulated in the Ising model. To this end, we have used the D-Wave 2000Q quantum annealer located at Los Alamos National Laboratory, as well as IBM's Poughkeepsie 20 qubit quantum computer available on the Oak Ridge National Laboratory IBM Q hub network together with the high-performance simulator, IBM Qiskit Aer Simulator~\cite{Qiskit}. However, our framework is modular and can easily be extended to utilize other novel quantum computing architectures as they become available. 

 The D-Wave 2000Q is the state-of-the-art quantum annealer at this time. It has up to 2048 qubits which are laid out in a special graph structure known as a Chimera graph. The Chimera graph is sparse, thus the device has sparse connectivity. Fully connected graphs as dense problems need to be embedded onto the device, which leads to the maximum size of 64 variables. We have used the embedding algorithm described in ~\cite{boothby2016fast} to calculate a complete embedding of the 64 variable problem. We found this embedding only once and reused it during our experiments. We utilized D-Wave's Solver API (SAPI) which is implemented in Python 2.7, to interact with the system. The D-Wave system is intrinsically a stochastic system, where solutions are sampled from a distribution corresponding to the lowest energy state. For each subproblem, the best solution out of 10,000 samples is returned. The annealing time for each call to the D-Wave system was set to 20 microseconds.

In order to solve problems formulated in the Ising model on IBM's Poughkeepsie quantum computer and simulator, we implemented QAOA using the  SBPLX~\cite{rowan1991functional} optimizer to find the optimal variational parameters. We allowed 2,000 iterations for SBPLX to find optimal parameters for QAOA. At each iteration, the circuit is executed 5,000 times (5,000 ``shots'') to obtain the statistic on the objective function. After the optimal parameters are found, the solution corresponding to best of 5,000 samples produced by running the ansatz with optimal parameters is returned. Due to the limitations of NISQ devices available in IBM Q hub network~\cite{Shaydulin2019}, we used the  RYRZ variational form~\cite{ibmqryrz-rs} (also known as a hardware-efficient ansatz) as the ansatz for our QAOA implementation. For the experiments run on IBM quantum device Poughkeepsie, we perform the variational parameter optimization on the simulator locally and run QAOA on the device via the IBM Q Experience cloud API. This is done due to the job queue limitations provided via the IBM Q Experience. However, we expect to be able to run QAOA variational parameter optimization fully on a device as more devices are becoming available on the cloud. We have used GNU Parallel ~\cite{tange_ole_2018_1146014} for the large-scale numerical experiments performed on the quantum simulator.

Considering the fact that solutions from the NISQ devices and simulator do not provide optimality guarantees,  we have also solved various subproblems formulated in the Ising model using the solver Gurobi \cite{optimization2014inc-rs} together with modeling package Pyomo \cite{hart2017pyomo}. The results using Gurobi as a solver for each subproblem are denoted as "Optimal" in our plots. Note that while each subproblem was solved and proven to be optimal for subproblem size 20, the same is not always true for subproblem size 64. For subproblem size 64 we occasionally observe non-zero gaps.

\paragraph{Instances}
A summary of the graphs used in the experiments together with their properties is presented in Table~\ref{tab:graph_prop}.
For the Graph Partitioning Problem, we evaluate ML-QLS on five graphs, four of which are drawn from The Graph Partitioning Archive~\cite{Soper2004} (\texttt{4elt}, \texttt{bcsstk30}, \texttt{cti} and \texttt{data}) and one from the set of hard to partition graphs
(\texttt{vsp\_msc10848\_300sep\_100in\_1Kout}, denoted in figures as \texttt{SSS12})~\cite{amg-sss12}. 
For the Modularity Maximization Problem, we evaluate ML-QLS on six graphs. The graphs 
\texttt{roadNet-PA-20k} and \texttt{opsahl-powergrid} are real-world networks from the KONECT dataset~\cite{kunegis2013konect_rs}. Graphs \texttt{msc23052} and \texttt{finan512-10k} are taken from the graph archive presented in ~\cite{Safro2011}. The graphs \texttt{finan512-10k} and \texttt{roadNet-PA-20k} are reduced to 10,000 and 20,000 nodes respectively by performing a breadth-first search from the median degree node. Note that due to the high diameter of these networks and their structure (portfolio optimization problem and road network), this preserves their structural properties.  \texttt{GirvanNewman} is a synthetic graph generated using the model introduced by Girvan and Newman (GN)~\cite{Girvan2002}. The graph  \texttt{lancichinetti1} is a synthetic graph generated using a generalization of the GN model that allows for heterogeneity in the distributions of node degree and community size, introduced by Lancichinetti et al.~\cite{Lancichinetti2008}. Table~\ref{tab:mod_synth} shows the parameters used to generate the synthetic graphs.

\begin{table*}[t]
\begin{center}
\begin{tabular}{ |c|c|c|c|c|c|c|c| } 
\hline 
 Network name & $|V|$ & $|E|$ & $d_{\mbox{avg}}$ & $d_{\mbox{max}}$  \\
 \hline
\texttt{SSS12} &	 21996 	& 1221028 &	 111.02 &	 722 \\
\texttt{4elt} &	 15606 &	 45878 &	 5.88 &	 10 \\
\texttt{bcsstk30} &	 28924 &	 1007284 &	 69.65 &	 218 \\
\texttt{cti} &	 16840 	& 48232 &	 5.73 & 	 6 \\
\texttt{data} &	 2851 &	 15093 	& 10.59 &	 17 \\
\texttt{roadNet-PA-20k} &	 20000 &	 26935 	& 2.69 &	 7 \\
\texttt{opsahl-powergrid} &	 4941 &	 6594 &	 2.67 &	 19 \\
\texttt{msc23052} &	 5722 &	 103391 &	 36.14 &	 125 \\
\texttt{finan512-10k} &	 10000 	& 28098 &	 5.62 &	 54 \\
 \hline
\end{tabular}
\caption{Properties of the networks used to evaluate ML-QLS. $d_{\mbox{avg}}$ is average degree, $d_{\mbox{max}}$ is maximum degree}
\label{tab:graph_prop}
\end{center}
\end{table*}

\begin{table*}[t]
\begin{center}
\begin{tabular}{ |c|c|c|c|c|c|c|c| } 
\hline 
 Network name & $|V|$ & $|E|$ & $d_{\mbox{avg}}$ & $d_{\mbox{max}}$ & $\gamma$ & $\beta$ & $\mu$ \\
 \hline
 \texttt{GirvanNewman} & 10,000 & 75,000 & 15.0 & 15 & 1 & 1 & 0.1 \\ 
 \texttt{lancichinetti1} & 10,000 & 76,133 & 15.22 & 50 & 2 & 1 & 0.1 \\ 
 \hline
\end{tabular}
\caption{Properties of synthetic networks used in the Modularity evaluation. $d_{\mbox{avg}}$ is average degree, $d_{\mbox{max}}$ is maximum degree, $\gamma$ is the  exponent for the degree distribution, $\beta$ is the exponent for the community size distribution and $\mu$ is the mixing parameter. For a detailed discussion of the parameters the reader is referred to Ref.~\cite{Lancichinetti2008}}
\label{tab:mod_synth}
\end{center}
\end{table*}

\paragraph{Experimental Setup} Our experiments are performed in order to compare the solutions from ML-QLS with those of high-quality classical solvers, and the best known results, if available. For the Graph Partitioning Problem, the results are compared to those produced by KaHIP~\cite{sandersschulz2013} which is a state-of-the-art multilevel Graph Partitioning solver. The best known results are taken at The Graph Partitioning Archive~\cite{Soper2004} where applicable. In order to make our approach more comparable to KaHIP, we follow the user guide~\cite{kahipguide}, and use the \texttt{kaffpaE} version of the solver with the option {\tt --mh\_enable\_kabapE} for high quality refinement for perfectly balanced parts. We use the option {\tt --preconfiguration=fast} to ensure results are compared with a single V-cycle. Our results (cut values) are normalized with either the best known value when applicable or by the smallest cut value found by any of the solvers used.

For the Modularity Maximization Problem, we compare our solutions using ML-QLS with two classical clustering methods, Asynchronous Fluid Communities~\cite{pares2017fluid} (implemented in NetworkX~\cite{hagberg2008-rs}) and Spectral Clustering~\cite{shi2000normalized, von2007tutorial-rs} (implemented in Scikit-learn~\cite{scikit-learn}). Note that even though these methods solve the same problem (namely, Community Detection or clustering), they do not explicitly maximize modularity. Therefore, it is unfair to directly compare the modularity of the solution produced by them to ML-QLS, which is explicitly maximizing modularity. However, they provide a useful baseline. Moreover, since the maximum possible modularity for at most 2 communities is 0.5, the best solutions found by all methods are no more than 1\%--10\% away from the optimal (see Table~\ref{tab:mod_best})

\begin{figure*}
    \centering
    \vspace{-1.5cm}
    \includegraphics[width=\textwidth, trim=0cm 4cm 0cm 0cm]{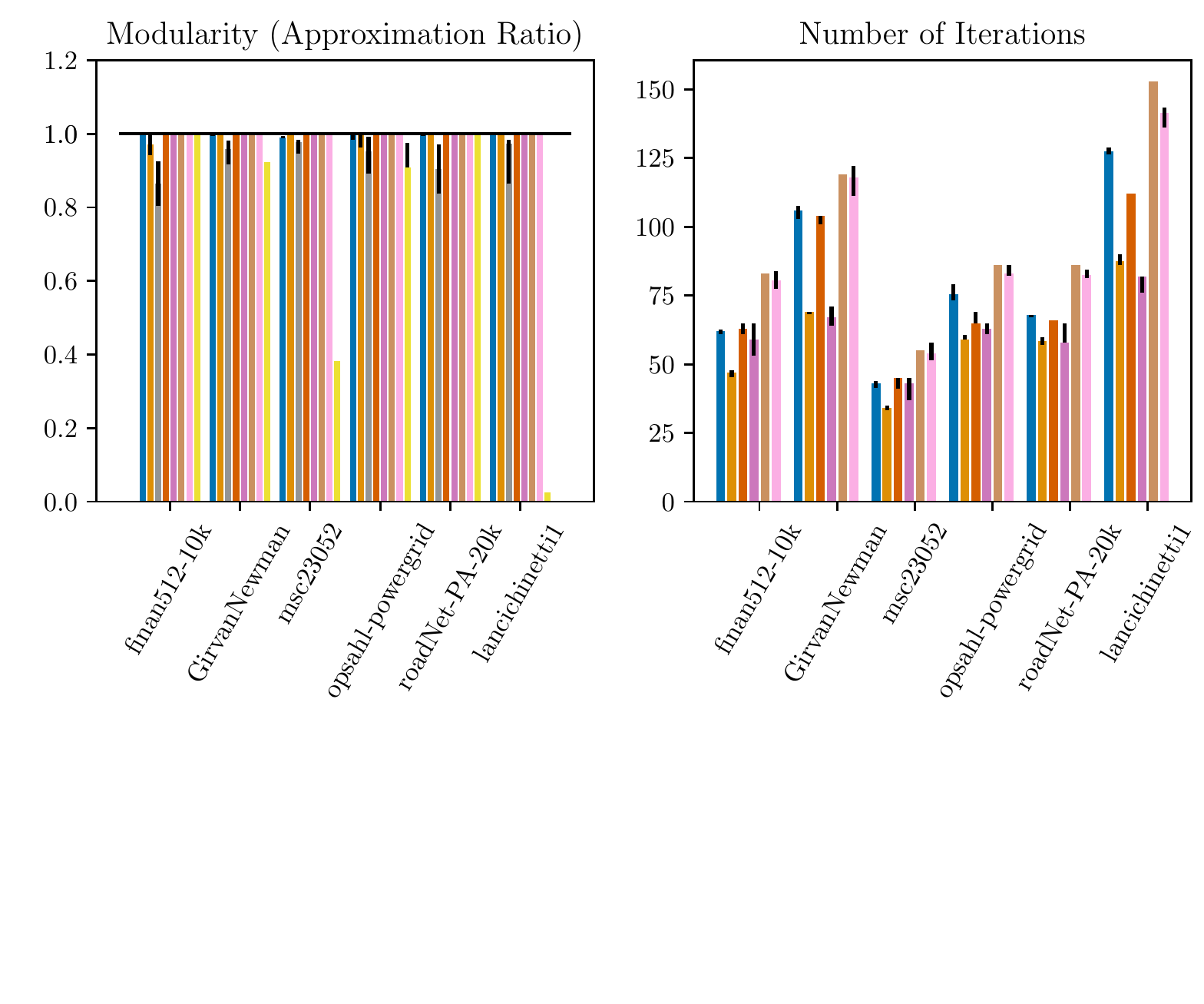}
    \includegraphics[width=\textwidth]{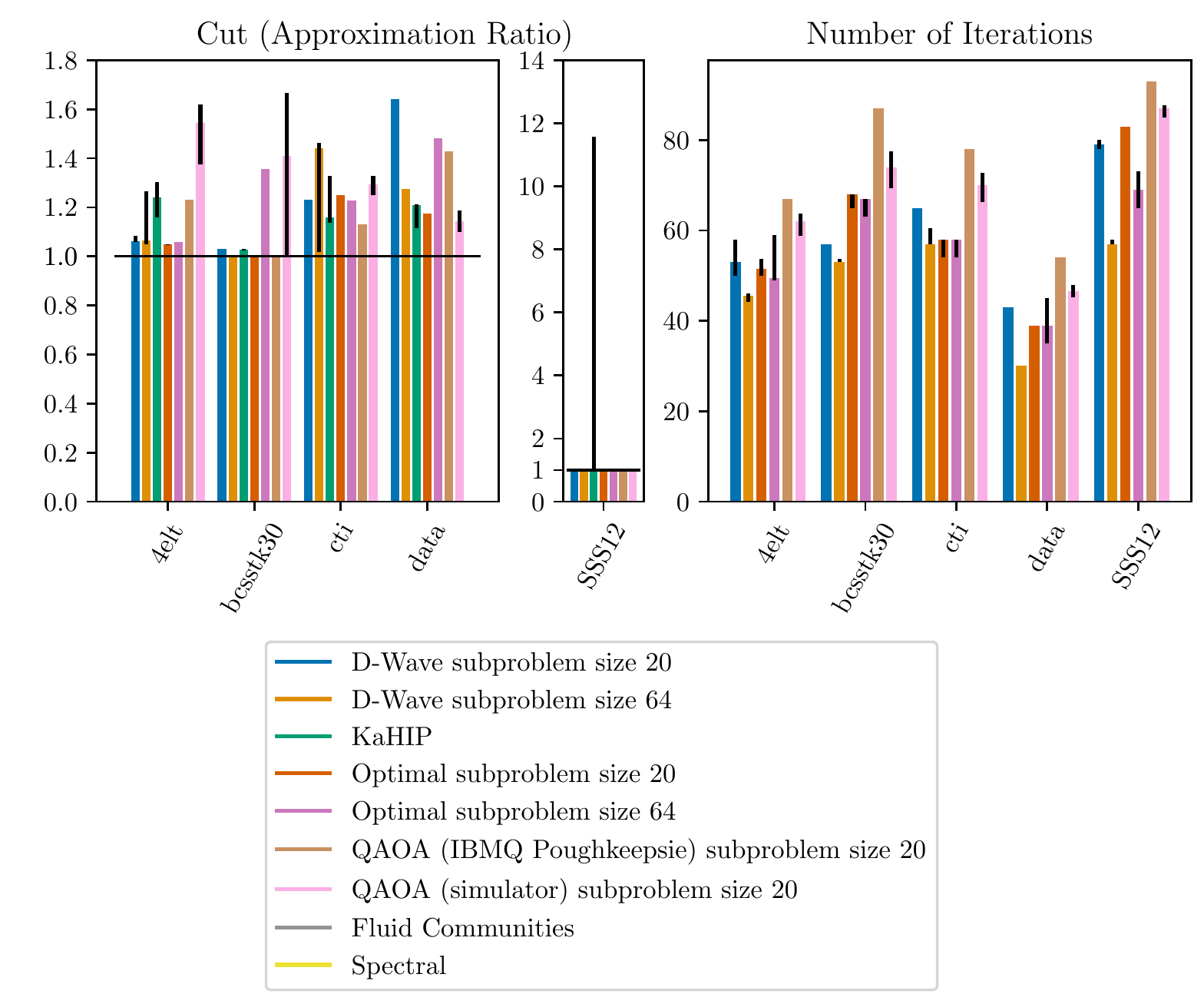}
    \caption{Quality of the solution and the number of iterations for all problems and solvers. The height of the bars is the median over 10 seeds. Error bars (black) are 25th and 75th percentiles. For the objective function (Cut or Modularity) all results are normalized by the best solution found by any solver (for Graph Partitioning this includes the best known cuts from The Graph Partitioning Archive~\cite{Soper2004}). Number of iterations is the number of calls to the subproblem solver (ML-QLS only).}
    \label{fig:barchart}
    \vspace{-0.25cm}
\end{figure*}

The experimental results are presented in Figure~\ref{fig:barchart}. We have made all raw result data available on Github~\cite{rawdata}. For each problem and method (except for QAOA on IBM Q Poughkeepsie quantum computer, labeled ``QAOA (IBMQ Poughkeepsie)'' in Figure~\ref{fig:barchart}), we perform ten runs of a single V-cycle with different seeds.  For ``QAOA (IBMQ Poughkeepsie)'', we perform just one run per each problem due to the limited access to quantum hardware.

\paragraph{Observations}

\begin{figure*}
\begin{center}
\begin{tabular}{ |c|c| } 
\hline 
 Network name & Best modularity\\
 \hline
 \texttt{finan512-10k}  & 0.499 \\
\texttt{GirvanNewman} & 0.459 \\
\texttt{lancichinetti1} & 0.452 \\
\texttt{msc23052}  & 0.499 \\
\texttt{opsahl-powergrid} & 0.497 \\
\texttt{roadNet-PA-20k} & 0.499 \\
 \hline
\end{tabular}
\caption{Highest modularity value found by all methods for a given problem. The highest possible modularity value for at most 2 communities is 0.5.}
\label{tab:mod_best}
\end{center}
\end{figure*}

 We observe that ML-QLS is capable of achieving results close to the best ones found by other solvers for all problems.  For Graph Partitioning, Figure~\ref{fig:barchart} shows significant variability in the quality of the solution across different solvers and problem instances. This effect is also observed for the state-of-the-art Graph Partitioning solver KaHIP, when run for a single V-cycle. This is partially due to the fact that we normalize the objectives to make them directly comparable. For example, for the graph {\tt 4elt} the best known cut value presented in The Graph Partitioning Archive~\cite{Soper2004} is 139. Therefore, an \emph{absolute} difference of 28 edges in cut obtained by a solver translates into a 20\% \emph{relative} difference presented in Figure~\ref{fig:barchart}. However, the same \emph{absolute} difference of 28 edges would translate into $\approx 0.44\%$ for the graph {\tt bcsstk30} (best known cut 6394). The graph \texttt{SSS12} is specifically designed to be hard for traditional Graph Partitioning frameworks~\cite{amg-sss12}. This explains the high variation in the performance of KaHIP on it.

It is worth noting that QAOA on the IBM quantum computers (see ``QAOA (IBM Q Poughkeepsie)'' in Figure~\ref{fig:barchart}) takes more iterations to converge to a solution compared to D-Wave. This is partially due to the fact that we perform the QAOA variational parameter optimization on the simulator and only run once with the optimized parameters on the device. As a result, the learned variational parameters do not include the noise profile of the device, limiting the quality of subproblem solutions. As devices become more easily available, we expect to be able to run full variational parameter optimization on the quantum hardware.

\begin{figure*}[tb]
    \centering
    \subfloat[Mean]{\includegraphics[width=0.5\textwidth]{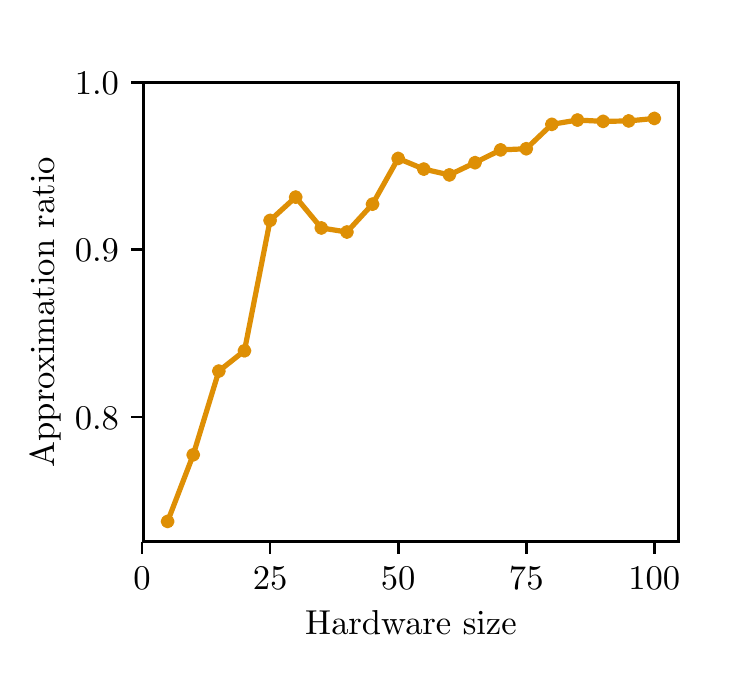}\label{fig:hwsize_scaling_mean}}
    \hfill
    \subfloat[Standard deviation]{\includegraphics[width=0.5\textwidth]{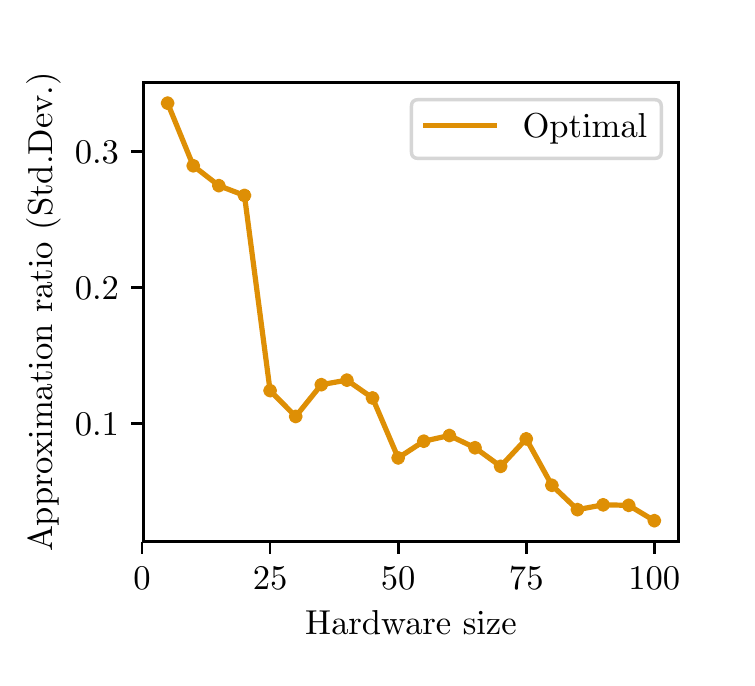}\label{fig:hwsize_scaling_std}}
    \caption{Modularity (Approximation ratio) as the function of the size of the subproblem (hardware size). The performance is projected using Gurobi as the subproblem solver. Fig.~\ref{fig:hwsize_scaling_mean} presents the mean approximation ratio averaged over the entire benchmark. Fig.~\ref{fig:hwsize_scaling_std} presents the standard deviation. As the hardware size increases, the quality of the solution found by ML-QLS improves.}
    \label{fig:hwsize_scaling}
\end{figure*}

\begin{figure*}
\includegraphics[width=0.5\textwidth]{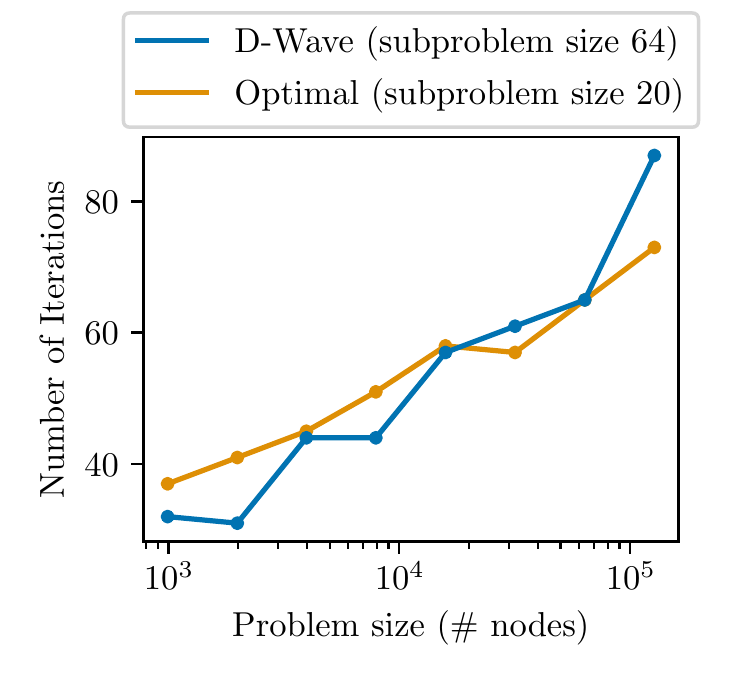}
\caption{The number of iterations (calls to optimizer) to solution for the modularity maximization problem as a function of problem size. The performance is projected using Gurobi as the subproblem solver with subproblem size 20 and allowing it to solve each subproblem to optimality, as well as using D-Wave as the subproblem solver with subproblem size 64. For Gurobi, the subproblem size was limited to 20 to guarantee that each subproblem is proven to be solved optimally.}
\label{fig:probsize_scaling}
\end{figure*}

To project the performance improvements for future hardware, we simulate the performance of ML-QLS as a function of hardware (subproblem) size shown in Figure~\ref{fig:hwsize_scaling}. As the subproblem size increases, the average quality of the solution found by ML-QLS improves and variation in results decreases. This shows that performance of ML-QLS can be improved as larger size quantum devices and better quantum optimization routines are developed. Note that this makes the assumption that the subproblem is solved to optimality or to a solution that is close to optimal. Evaluating the scaling of quantum optimization algorithms used for solving the subproblems falls outside of the scope of this paper (we provide an overview of relevant recent results in Sec.~\ref{sec:qopt_scale}).

\subsection{Scaling and running time estimates}\label{sec:exp_scale}

The proposed ML-QLS approach is based on the traditional multilevel methods for graph partitioning and graph clustering and therefore a lot of scaling and running time considerations are shared between the two family of methods. Concretely, in our implementation we use the coarsening available in KaHIP graph partitioning package~\cite{sandersschulz2013}, making the running time and the scaling of the \emph{coarsening} stage of our method and KaHIP equal. The running time of solving the problem on the coarsest level does not scale with problem size, as the size of the coarsest level is fixed to be equal to the hardware size. Therefore in this section we will focus on the analysis of the \emph{refinement} stage.

To evaluate how the proposed approach scales with the problem size, we construct a series of graphs with the number of nodes ranging from $1,000$ to $128,000$. The graphs are constructed in the same way as the graph \texttt{roadNet-PA-20k}, namely by performing a breadth-first search from the median degree node of \texttt{roadNet-PA}~\cite{KONECT13} and including nodes until the desired size is reached. We fix the subproblem size to 20 for Gurobi and to 64 for D-Wave. The results are presented in Fig.~\ref{fig:probsize_scaling}. We note that the number of iterations scales roughly logarithmically with the problem size, as we observe roughly constant number of iterations per uncoarsening level. As we do not constrain the number of refinement iterations, this indicates that the solution projected from coarser levels is of high quality and does not need to be significantly changed at the refinement stage, indicating that the coarsening has successfully constructed a multilevel hierarchy that preserves problem structure. The success of coarsening in this example might be due to the simplicity of the problem structure (planar graph of a road network). However, in all major multilevel solvers the number of refinement calls is artificially limited, preserving the scaling (see Sec.~\ref{sec:multilevel_scale} for an in-depth discussion).

Due to the high overhead of queuing system and remote execution on cloud quantum computers, the actual observed running time in our experiments is not representative of the algorithm performance. To give the reader a sense of the running time of our approach, we use the number of iterations to solution presented in Fig.~\ref{fig:barchart} and \ref{fig:probsize_scaling} to compute rough estimates of the running time of the algorithm. To do that, we briefly discuss the running time of QAOA and quantum annealing for our problems. We observe that as the running time of coarsening is approximately 1 second for the problems in our benchmark, the overall running time is dominated by refinement. Same is true of this class of multilevel algorithms in general (see Sec.~\ref{sec:multilevel_scale}).

To understand the running time of QAOA, we need to understand two of its components: the ``training time'' (time to find good QAOA parameters) and the ``execution time'' (time to sample from QPU with optimal parameters). In this work, we train our variational quantum optimizer (which can be considered a version of QAOA as described in Section~\ref{sec:background:qopt}) purely classically in simulation. This results in running time of $\approx300$ seconds for 20 qubits and 2000 iterations of outer-loop optimization using quantum simulator Qiskit Aer and these runtimes can be improved using other more efficient state-vector simulators~\cite{wu2019full}. However, this is approach becomes unrealistic as problem size grows. While novel tensor-network based approaches allow purely classical training of certain class of QAOA instances~\cite{Streif2020}, they are not applicable to the problems considered in this paper due to the full connectivity of the problem. Therefore there are currently no viable approaches to classically training QAOA for larger modularity and graph partitioning problems. Running time for larger problems can be estimated by taking the ``execution time'' and multiplying it by number of iterations of outer-loop optimization. At the time of writing, IBM Quantum Experience did not allow us to directly measure the execution time of the quantum circuit with optimal parameters. Therefore we use the running time reports available in the literature, from which we estimate the walltime of $\approx 1s$ for 5,000 shots~\cite{karalekas2020optimizedforvariational,Arute2019}. Combined with the purely classical training time reported above, for a problem that takes 60 iterations to solve (e.g. modularity maximization for \texttt{4elt}), this gives projected running time of $\approx 30$ minutes. The actual running time observed in experiment has been larger due to the additional overhead introduced by the queue system (up to hours per iteration, depending on queue state).

For Quantum Annealing on the D-Wave system, the overall running time spent on the device is directly proportional to the number of iterations of the algorithm as reported in Figure \ref{fig:probsize_scaling}. For a fixed annealing time, and number of samples requested, the running time on the device is approximately constant per iteration. For an annealing time of 20 microseconds, together with 1000 samples, in our experiments we observed an overall running time of approximately 5 seconds for the 15 iterations when applied to the graph of 1000, while taking about 15 seconds when applied to the graph of approximately 127,000 nodes.

%

\section{Open Problems and Discussion}\label{sec:conclusion}
Revising (un)coarsening operators in anticipation of the new class of high-quality  refinement solvers is the first major open problem. The majority of multilevel algorithms for combinatorial optimization problems are inspired by the idea of "thinking globally while acting locally".
However, there is a crucial difference between  these algorithms for combinatorial problems and such methods as multigrid for continuous problems or multiscale PDE-based optimization. 
In multigrid (e.g., for systems of linear equations), a relaxation at the uncoarsening stage is convergent \cite{brandt:review01}, and in most cases assumes an optimal solution (up to some tolerance) for a subset of relaxed variables given other variables are invariant (i.e., a fixed solution for those variables that are not in the optimized subset). Examples include easily parallelizable Jacobi relaxation, as well as  hard to parallelize Gauss-Seidel relaxation in which most variables are typically optimized sequentially, and many more. Both coarsening and uncoarsening operators (also known as the restriction, and prolongation in multigrid) assume this convergence which in the end provides guarantees for the entire multilevel framework. However, for the \emph{combinatorial} multilevel solvers, the integer variables make this assumption practically impossible,  even for subproblems containing tens of variables optimized simultaneously. With the development of less noisy quantum devices, we can assume that in our hands will be extremely fast heuristics to produce nearly (if hypothetically not completely) optimal solutions for combinatorial optimization problems of up to several hundreds of variables. In order to use the multilevel paradigm correctly, there will be a critical need to revise  (un)coarsening operators that take this feature into account because (to the best of our knowledge) all existing versions of coarsening operators do not consider optimality of the refinement.  Moreover, most existing multilevel frameworks exhibit more emphasis on computational speedup rather than on the quality of the solution to better approximate the fine problem.

The second problem is not unique to multilevel methods but to most decomposition based approaches. Even if quantum devices become fully developed and become more accessible for the broad scientific community, they will still remain more expensive than regular CPU based devices. The decomposition approaches split the problem into many small local subproblems, while multilevel methods may need even more of them because solving subproblems is required at all levels of coarseness. Thus, there is a critical need in developing an extremely fast routing classifier for a subproblem that will decide whether solving a particular subproblem  on the NISQ device will be beneficial in comparison to the CPU.

The third consideration is the sparsity of the problem. The methods outlined in this work are evaluated on sparse problems, and we expect them to perform well only under the assumption on sparsity of the problem. We make this assumption because the standard benchmarks for the problems we consider are sparse (see, for example, The Graph Partitioning Archive~\cite{Soper2004} and DIMACS Graph Partitioning and Graph Clustering Challenge~\cite{sanders2014benchmarking,bader2013graph}). Development of decomposition methods specifically targeting dense problems is an interesting future direction. Typically, in existing  multilevel solvers, the fact of density at some level serves as a stopping criterion, i.e., when the problem is originally dense, the multilevel hierarchy construction is terminated which is clearly insufficient thing to do for many distributions of accumulated edge weights. In other words, this problem exists not only for multilevel quantum but for classical solvers as well. In multilevel algorithms, the problem density requires a different approach such as sparsification \cite{safro:spars}.


\section{Conclusion} Current Noisy Intermediate-Scale Quantum (NISQ) devices are limited in the number of qubits and can therefore only be used to directly solve combinatorial optimization problems that exhibit a limited number of variables. In order to overcome this limitation, in this work we have proposed the multilevel computational framework for solving large-scale combinatorial problems on NISQ devices. We demonstrate this approach on two well-known combinatorial optimization problems, the Graph Partitioning Problem, and the Community Detection Problem, and perform experiments on the 20 qubit IBM gate-model quantum computer, and the 2048 qubit D-Wave 2000Q quantum annealer. In order to implement an efficient iterative refinement scheme using the NISQ devices, we have developed novel techniques for efficiently formulating and evaluating sub-QUBOs without explicitly constructing the entire QUBO of the large-scale problem, which in many cases can be a dense matrix that makes it computationally expensive to store and process. In our experiments, for the Graph Partitioning Problem, five graphs were chosen such that the smallest graph had 2851 nodes while the largest had 28924 nodes, while for the Community Detection Problem, the smallest graph had 4941 nodes and largest had 10,000 nodes. For both problems, for comparison purposes, we run one V-cycle of the multilevel framework with the different NISQ devices multiple times and compared the results to the state-of-art methods. Our experimental results give comparable results to the state-of-the-art methods and for some cases we were able to get the best-known results.  This work therefore provides an important stepping stone to demonstrating practical Quantum Advantage. As the capabilities of NISQ devices increase, we are hopeful that similar methods can provide a path to adoption of quantum computers for a variety of business~\cite{dwavefinance} and scientific applications.


\section*{Acknowledgments}
The authors thank anonymous referees whose valuable comments helped to  improve this work. This work was supported in part with funding from the Defense Advanced Research Projects Agency (DARPA). The views, opinions and/or findings expressed are those of the author and should not be interpreted as representing the official views or policies of the Department of Defense or the U.S. Government. 
This work was supported in part by NSF award \#1522751. High-performance computing resources at Clemson University were supported by NSF award MRI \#1725573. This research used resources of the Oak Ridge Leadership Computing Facility, which is a DOE Office of Science User Facility supported under Contract DE-AC05-00OR22725. This research also used the resources of the Argonne Leadership Computing Facility, which is DOE Office of Science User Facility supported under Contract DE-AC02-06CH11357. Yuri Alexeev and Ruslan Shaydulin were supported by the DOE Office of Science. The authors would also like to acknowledge the NNSA’s Advanced Simulation and Computing (ASC) program at Los Alamos National Laboratory (LANL) for use of their Ising D-Wave 2000Q quantum computing resource. LANL is operated by Triad National Security, LLC, for the National Nuclear Security Administration of U.S. Department of Energy (Contract No. 89233218NCA000001). Susan Mniszewski and Christian Negre were supported by the ASC program at LANL. Assigned: Los Alamos Unclassified Report 19-30113.

	\bibliographystyle{ACM-Reference-Format}
	\bibliography{bib,ruslan,fullbib}


\begin{thebibliography}{105}


\ifx \showCODEN    \undefined \def \showCODEN     #1{\unskip}     \fi
\ifx \showDOI      \undefined \def \showDOI       #1{#1}\fi
\ifx \showISBNx    \undefined \def \showISBNx     #1{\unskip}     \fi
\ifx \showISBNxiii \undefined \def \showISBNxiii  #1{\unskip}     \fi
\ifx \showISSN     \undefined \def \showISSN      #1{\unskip}     \fi
\ifx \showLCCN     \undefined \def \showLCCN      #1{\unskip}     \fi
\ifx \shownote     \undefined \def \shownote      #1{#1}          \fi
\ifx \showarticletitle \undefined \def \showarticletitle #1{#1}   \fi
\ifx \showURL      \undefined \def \showURL       {\relax}        \fi
\providecommand\bibfield[2]{#2}
\providecommand\bibinfo[2]{#2}
\providecommand\natexlab[1]{#1}
\providecommand\showeprint[2][]{arXiv:#2}

\bibitem[\protect\citeauthoryear{??}{cod}{[n. d.]}]%
        {code}
 \bibinfo{year}{[n. d.]}\natexlab{}.
\newblock
\newblock
\urldef\tempurl%
\url{https://github.com/rsln-s/ml_qls}
\showURL{%
\tempurl}


\bibitem[\protect\citeauthoryear{??}{raw}{[n. d.]}]%
        {rawdata}
 \bibinfo{year}{[n. d.]}\natexlab{}.
\newblock
\newblock
\urldef\tempurl%
\url{https://github.com/rsln-s/ml_qls/tree/bc376276ba684460aeccaa371b4fc38003139e34/multilevel/data/results_csv}
\showURL{%
\tempurl}


\bibitem[\protect\citeauthoryear{??}{ibm}{[n. d.]}]%
        {ibmqryrz-rs}
 \bibinfo{year}{[n. d.]}\natexlab{}.
\newblock \bibinfo{title}{{IBM QISKit Aqua: Variatinal forms}}.
\newblock
  \bibinfo{howpublished}{\url{https://github.com/Qiskit/qiskit-aqua/blob/master/qiskit/aqua/components/variational_forms/ryrz.py}}.
\newblock
\newblock
\shownote{[Online; accessed July 16, 2019].}


\bibitem[\protect\citeauthoryear{??}{kah}{[n. d.]}]%
        {kahipguide}
 \bibinfo{year}{[n. d.]}\natexlab{}.
\newblock \bibinfo{title}{{KaHIP v2.10 – Karlsruhe High Quality Partitioning
  User Guide}}.
\newblock
  \bibinfo{howpublished}{\url{http://algo2.iti.kit.edu/schulz/software_releases/kahipv2.10.pdf}}.
\newblock
\newblock
\shownote{[Online; accessed July 31, 2019].}


\bibitem[\protect\citeauthoryear{??}{iar}{[n. d.]}]%
        {iarpaqeo}
 \bibinfo{year}{[n. d.]}\natexlab{}.
\newblock \bibinfo{title}{{Quantum Enhanced Optimization (QEO)}}.
\newblock
  \bibinfo{howpublished}{\url{https://www.iarpa.gov/index.php/research-programs/qeo}}.
\newblock
\newblock
\shownote{[Online; accessed September 25, 2018].}


\bibitem[\protect\citeauthoryear{Albash and Lidar}{Albash and Lidar}{2018a}]%
        {albash2a018adiabatic-rs}
\bibfield{author}{\bibinfo{person}{Tameem Albash} {and}
  \bibinfo{person}{Daniel~A Lidar}.} \bibinfo{year}{2018}\natexlab{a}.
\newblock \showarticletitle{Adiabatic quantum computation}.
\newblock \bibinfo{journal}{\emph{Reviews of Modern Physics}}
  \bibinfo{volume}{90}, \bibinfo{number}{1} (\bibinfo{year}{2018}),
  \bibinfo{pages}{015002}.
\newblock


\bibitem[\protect\citeauthoryear{Albash and Lidar}{Albash and Lidar}{2018b}]%
        {Albash2018-rs}
\bibfield{author}{\bibinfo{person}{Tameem Albash} {and}
  \bibinfo{person}{Daniel~A. Lidar}.} \bibinfo{year}{2018}\natexlab{b}.
\newblock \showarticletitle{Demonstration of a Scaling Advantage for a Quantum
  Annealer over Simulated Annealing}.
\newblock \bibinfo{journal}{\emph{Physical Review X}} \bibinfo{volume}{8},
  \bibinfo{number}{3} (\bibinfo{date}{July} \bibinfo{year}{2018}).
\newblock
\urldef\tempurl%
\url{https://doi.org/10.1103/physrevx.8.031016}
\showDOI{\tempurl}


\bibitem[\protect\citeauthoryear{Aleksandrowicz, Alexander, Barkoutsos, Bello,
  Ben-Haim, et~al\mbox{.}}{Aleksandrowicz et~al\mbox{.}}{2019}]%
        {Qiskit}
\bibfield{author}{\bibinfo{person}{Gadi Aleksandrowicz},
  \bibinfo{person}{Thomas Alexander}, \bibinfo{person}{Panagiotis Barkoutsos},
  \bibinfo{person}{Luciano Bello}, \bibinfo{person}{Yael Ben-Haim},
  {et~al\mbox{.}}} \bibinfo{year}{2019}\natexlab{}.
\newblock \bibinfo{title}{Qiskit: An Open-source Framework for Quantum
  Computing}.
\newblock
\newblock
\urldef\tempurl%
\url{https://doi.org/10.5281/zenodo.2562110}
\showDOI{\tempurl}


\bibitem[\protect\citeauthoryear{Arute, Arya, Babbush, Bacon, Bardin,
  et~al\mbox{.}}{Arute et~al\mbox{.}}{2019}]%
        {Arute2019}
\bibfield{author}{\bibinfo{person}{Frank Arute}, \bibinfo{person}{Kunal Arya},
  \bibinfo{person}{Ryan Babbush}, \bibinfo{person}{Dave Bacon},
  \bibinfo{person}{Joseph~C. Bardin}, {et~al\mbox{.}}}
  \bibinfo{year}{2019}\natexlab{}.
\newblock \showarticletitle{Quantum supremacy using a programmable
  superconducting processor}.
\newblock \bibinfo{journal}{\emph{Nature}} \bibinfo{volume}{574},
  \bibinfo{number}{7779} (\bibinfo{date}{Oct.} \bibinfo{year}{2019}),
  \bibinfo{pages}{505--510}.
\newblock
\urldef\tempurl%
\url{https://doi.org/10.1038/s41586-019-1666-5}
\showDOI{\tempurl}


\bibitem[\protect\citeauthoryear{Bader, Meyerhenke, Sanders, and Wagner}{Bader
  et~al\mbox{.}}{2013}]%
        {bader2013graph}
\bibfield{author}{\bibinfo{person}{David~A Bader}, \bibinfo{person}{Henning
  Meyerhenke}, \bibinfo{person}{Peter Sanders}, {and} \bibinfo{person}{Dorothea
  Wagner}.} \bibinfo{year}{2013}\natexlab{}.
\newblock \bibinfo{booktitle}{\emph{Graph partitioning and graph clustering}}.
  Vol.~\bibinfo{volume}{588}.
\newblock \bibinfo{publisher}{American Mathematical Society Providence, RI}.
\newblock


\bibitem[\protect\citeauthoryear{Booth, Reinhardt, and Roy}{Booth
  et~al\mbox{.}}{2017}]%
        {booth2017partitioning}
\bibfield{author}{\bibinfo{person}{Michael Booth}, \bibinfo{person}{Steven~P
  Reinhardt}, {and} \bibinfo{person}{Aidan Roy}.}
  \bibinfo{year}{2017}\natexlab{}.
\newblock \showarticletitle{Partitioning optimization problems for hybrid
  classical}.
\newblock \bibinfo{journal}{\emph{quantum execution. Technical Report}}
  (\bibinfo{year}{2017}), \bibinfo{pages}{01--09}.
\newblock


\bibitem[\protect\citeauthoryear{Boothby, King, and Roy}{Boothby
  et~al\mbox{.}}{2016}]%
        {boothby2016fast}
\bibfield{author}{\bibinfo{person}{Tomas Boothby}, \bibinfo{person}{Andrew~D
  King}, {and} \bibinfo{person}{Aidan Roy}.} \bibinfo{year}{2016}\natexlab{}.
\newblock \showarticletitle{Fast clique minor generation in Chimera qubit
  connectivity graphs}.
\newblock \bibinfo{journal}{\emph{Quantum Information Processing}}
  \bibinfo{volume}{15}, \bibinfo{number}{1} (\bibinfo{year}{2016}),
  \bibinfo{pages}{495--508}.
\newblock


\bibitem[\protect\citeauthoryear{Brandao, Broughton, Farhi, Gutmann, and
  Neven}{Brandao et~al\mbox{.}}{2018}]%
        {brandao2018fixed}
\bibfield{author}{\bibinfo{person}{Fernando~GSL Brandao},
  \bibinfo{person}{Michael Broughton}, \bibinfo{person}{Edward Farhi},
  \bibinfo{person}{Sam Gutmann}, {and} \bibinfo{person}{Hartmut Neven}.}
  \bibinfo{year}{2018}\natexlab{}.
\newblock \showarticletitle{For Fixed Control Parameters the Quantum
  Approximate Optimization Algorithm's Objective Function Value Concentrates
  for Typical Instances}.
\newblock \bibinfo{journal}{\emph{arXiv preprint arXiv:1812.04170}}
  (\bibinfo{year}{2018}).
\newblock


\bibitem[\protect\citeauthoryear{Brandt}{Brandt}{2002}]%
        {brandt:review01}
\bibfield{author}{\bibinfo{person}{A. Brandt}.}
  \bibinfo{year}{2002}\natexlab{}.
\newblock \showarticletitle{{Multiscale Scientific Computation: Review 2001}}.
\newblock In \bibinfo{booktitle}{\emph{{Multiscale and Multiresolution
  Methods}}}. \bibinfo{series}{LNCSE}, Vol.~\bibinfo{volume}{20}.
  \bibinfo{publisher}{Springer}, \bibinfo{pages}{3--95}.
\newblock
\showISBNx{978-3-540-42420-8}
\urldef\tempurl%
\url{http://dx.doi.org/10.1007/978-3-642-56205-1_1}
\showURL{%
\tempurl}


\bibitem[\protect\citeauthoryear{Brandt and Ron}{Brandt and Ron}{2003}]%
        {vlsicad}
\bibfield{author}{\bibinfo{person}{A. Brandt} {and} \bibinfo{person}{D. Ron}.}
  \bibinfo{year}{2003}\natexlab{}.
\newblock \showarticletitle{Chapter 1 : Multigrid solvers and multilevel
  optimization strategies}.
\newblock In \bibinfo{booktitle}{\emph{Multilevel Optimization and VLSICAD}},
  \bibfield{editor}{\bibinfo{person}{J.~Cong} {and} \bibinfo{person}{J.~R.
  Shinnerl}} (Eds.). \bibinfo{publisher}{Kluwer}.
\newblock


\bibitem[\protect\citeauthoryear{Bulu{\c{c}}, Meyerhenke, Safro, Sanders, and
  Schulz}{Bulu{\c{c}} et~al\mbox{.}}{2016}]%
        {bulucc2016recent}
\bibfield{author}{\bibinfo{person}{Ayd{\i}n Bulu{\c{c}}},
  \bibinfo{person}{Henning Meyerhenke}, \bibinfo{person}{Ilya Safro},
  \bibinfo{person}{Peter Sanders}, {and} \bibinfo{person}{Christian Schulz}.}
  \bibinfo{year}{2016}\natexlab{}.
\newblock \showarticletitle{Recent advances in graph partitioning}.
\newblock In \bibinfo{booktitle}{\emph{Algorithm Engineering: Selected Results
  and Surveys. LNCS 9220, Springer-Verlag}}. \bibinfo{publisher}{Springer},
  \bibinfo{pages}{117--158}.
\newblock


\bibitem[\protect\citeauthoryear{Cerezo, Sone, Volkoff, Cincio, and
  Coles}{Cerezo et~al\mbox{.}}{2020}]%
        {cerezo2020costfunctiondependent}
\bibfield{author}{\bibinfo{person}{M. Cerezo}, \bibinfo{person}{Akira Sone},
  \bibinfo{person}{Tyler Volkoff}, \bibinfo{person}{Lukasz Cincio}, {and}
  \bibinfo{person}{Patrick~J. Coles}.} \bibinfo{year}{2020}\natexlab{}.
\newblock \bibinfo{title}{Cost-Function-Dependent Barren Plateaus in Shallow
  Quantum Neural Networks}.
\newblock
\newblock
\showeprint{arXiv:2001.00550}


\bibitem[\protect\citeauthoryear{Cong and Shinnerl}{Cong and Shinnerl}{2003}]%
        {Cong2003}
\bibfield{editor}{\bibinfo{person}{J. Cong} {and} \bibinfo{person}{J.~R.
  Shinnerl}} (Eds.). \bibinfo{year}{2003}\natexlab{}.
\newblock \bibinfo{booktitle}{\emph{Multilevel Optimization and VLSICAD}}.
\newblock \bibinfo{publisher}{Kluwer}.
\newblock
\showISBNx{1-4020-1081-8}


\bibitem[\protect\citeauthoryear{Crooks}{Crooks}{2018}]%
        {crooks2018performance}
\bibfield{author}{\bibinfo{person}{Gavin~E Crooks}.}
  \bibinfo{year}{2018}\natexlab{}.
\newblock \showarticletitle{Performance of the quantum approximate optimization
  algorithm on the maximum cut problem}.
\newblock \bibinfo{journal}{\emph{arXiv:1811.08419}} (\bibinfo{year}{2018}).
\newblock


\bibitem[\protect\citeauthoryear{{D-Wave Systems Inc.}}{{D-Wave Systems
  Inc.}}{2018}]%
        {dwave2018}
\bibfield{author}{\bibinfo{person}{{D-Wave Systems Inc.}}}
  \bibinfo{year}{2018}\natexlab{}.
\newblock \showarticletitle{Introduction to the {D-Wave} Quantum Hardware}.
\newblock  (\bibinfo{year}{2018}).
\newblock
\urldef\tempurl%
\url{www.dwavesys.com/tutorials/background-reading-series/introduction-d-wave-quantum-hardware}
\showURL{%
\tempurl}


\bibitem[\protect\citeauthoryear{{D-Wave Systems Inc.}}{{D-Wave Systems
  Inc.}}{2019}]%
        {Dwave2019time}
\bibfield{author}{\bibinfo{person}{{D-Wave Systems Inc.}}}
  \bibinfo{year}{2019}\natexlab{}.
\newblock \showarticletitle{Measuring Computation Time on D-Wave Systems}.
\newblock \bibinfo{journal}{\emph{D-Wave User Manual 09-1107A-M}}
  (\bibinfo{year}{2019}).
\newblock
\urldef\tempurl%
\url{https://docs.dwavesys.com/docs/latest/doc_timing.html}
\showURL{%
\tempurl}


\bibitem[\protect\citeauthoryear{Davis, Hager, Kolodziej, and Yeralan}{Davis
  et~al\mbox{.}}{2019}]%
        {davis2019algorithm}
\bibfield{author}{\bibinfo{person}{Timothy~A Davis}, \bibinfo{person}{William~W
  Hager}, \bibinfo{person}{Scott~P Kolodziej}, {and} \bibinfo{person}{S~Nuri
  Yeralan}.} \bibinfo{year}{2019}\natexlab{}.
\newblock \showarticletitle{Algorithm XXX: Mongoose, a graph coarsening and
  partitioning library}.
\newblock \bibinfo{journal}{\emph{ACM Trans. Math. Software}}
  (\bibinfo{year}{2019}).
\newblock


\bibitem[\protect\citeauthoryear{Denchev, Boixo, Isakov, Ding, Babbush,
  et~al\mbox{.}}{Denchev et~al\mbox{.}}{2016}]%
        {Denchev2016}
\bibfield{author}{\bibinfo{person}{Vasil~S. Denchev}, \bibinfo{person}{Sergio
  Boixo}, \bibinfo{person}{Sergei~V. Isakov}, \bibinfo{person}{Nan Ding},
  \bibinfo{person}{Ryan Babbush}, {et~al\mbox{.}}}
  \bibinfo{year}{2016}\natexlab{}.
\newblock \showarticletitle{What is the Computational Value of Finite-Range
  Tunneling?}
\newblock \bibinfo{journal}{\emph{Physical Review X}} \bibinfo{volume}{6},
  \bibinfo{number}{3} (\bibinfo{date}{Aug.} \bibinfo{year}{2016}).
\newblock
\urldef\tempurl%
\url{https://doi.org/10.1103/physrevx.6.031015}
\showDOI{\tempurl}


\bibitem[\protect\citeauthoryear{Ding, Lamata, Martín-Guerrero, Lizaso, Mugel,
  et~al\mbox{.}}{Ding et~al\mbox{.}}{2019}]%
        {dwavefinance}
\bibfield{author}{\bibinfo{person}{Yongcheng Ding}, \bibinfo{person}{Lucas
  Lamata}, \bibinfo{person}{José~D. Martín-Guerrero},
  \bibinfo{person}{Enrique Lizaso}, \bibinfo{person}{Samuel Mugel},
  {et~al\mbox{.}}} \bibinfo{year}{2019}\natexlab{}.
\newblock \bibinfo{title}{Towards Prediction of Financial Crashes with a D-Wave
  Quantum Computer}.
\newblock
\newblock
\showeprint{arXiv:1904.05808}


\bibitem[\protect\citeauthoryear{Elgart and Hagedorn}{Elgart and
  Hagedorn}{2012}]%
        {Elgart2012}
\bibfield{author}{\bibinfo{person}{Alexander Elgart} {and}
  \bibinfo{person}{George~A. Hagedorn}.} \bibinfo{year}{2012}\natexlab{}.
\newblock \showarticletitle{A note on the switching adiabatic theorem}.
\newblock \bibinfo{journal}{\emph{J. Math. Phys.}} \bibinfo{volume}{53},
  \bibinfo{number}{10} (\bibinfo{date}{Oct.} \bibinfo{year}{2012}),
  \bibinfo{pages}{102202}.
\newblock
\urldef\tempurl%
\url{https://doi.org/10.1063/1.4748968}
\showDOI{\tempurl}


\bibitem[\protect\citeauthoryear{Farhi, Goldstone, Gutmann, Lapan, Lundgren,
  et~al\mbox{.}}{Farhi et~al\mbox{.}}{2001}]%
        {quant-ph/0104129}
\bibfield{author}{\bibinfo{person}{Edward Farhi}, \bibinfo{person}{Jeffrey
  Goldstone}, \bibinfo{person}{Sam Gutmann}, \bibinfo{person}{Joshua Lapan},
  \bibinfo{person}{Andrew Lundgren}, {et~al\mbox{.}}}
  \bibinfo{year}{2001}\natexlab{}.
\newblock \showarticletitle{A Quantum Adiabatic Evolution Algorithm Applied to
  Random Instances of an NP-Complete Problem}.
\newblock  (\bibinfo{year}{2001}).
\newblock
\urldef\tempurl%
\url{https://doi.org/10.1126/science.1057726}
\showDOI{\tempurl}
\showeprint[arXiv]{quant-ph/0104129}


\bibitem[\protect\citeauthoryear{Farhi, Goldstone, Gutmann, and Sipser}{Farhi
  et~al\mbox{.}}{2000}]%
        {quant-ph/0001106}
\bibfield{author}{\bibinfo{person}{Edward Farhi}, \bibinfo{person}{Jeffrey
  Goldstone}, \bibinfo{person}{Sam Gutmann}, {and} \bibinfo{person}{Michael
  Sipser}.} \bibinfo{year}{2000}\natexlab{}.
\newblock \bibinfo{title}{Quantum Computation by Adiabatic Evolution}.
\newblock
\newblock
\showeprint[arXiv]{quant-ph/0001106}


\bibitem[\protect\citeauthoryear{Garcia-Saez and Riu}{Garcia-Saez and
  Riu}{2019}]%
        {garcia2019quantum}
\bibfield{author}{\bibinfo{person}{Artur Garcia-Saez} {and}
  \bibinfo{person}{Jordi Riu}.} \bibinfo{year}{2019}\natexlab{}.
\newblock \showarticletitle{Quantum Observables for continuous control of the
  Quantum Approximate Optimization Algorithm via Reinforcement Learning}.
\newblock \bibinfo{journal}{\emph{arXiv preprint arXiv:1911.09682}}
  (\bibinfo{year}{2019}).
\newblock


\bibitem[\protect\citeauthoryear{Gelman and Mandel}{Gelman and Mandel}{1990}]%
        {gelman1990multilevel}
\bibfield{author}{\bibinfo{person}{E Gelman} {and} \bibinfo{person}{J Mandel}.}
  \bibinfo{year}{1990}\natexlab{}.
\newblock \showarticletitle{On multilevel iterative methods for optimization
  problems}.
\newblock \bibinfo{journal}{\emph{Mathematical Programming}}
  \bibinfo{volume}{48}, \bibinfo{number}{1-3} (\bibinfo{year}{1990}),
  \bibinfo{pages}{1--17}.
\newblock


\bibitem[\protect\citeauthoryear{Girvan and Newman}{Girvan and Newman}{2002}]%
        {Girvan2002}
\bibfield{author}{\bibinfo{person}{M. Girvan} {and} \bibinfo{person}{M.~E.~J.
  Newman}.} \bibinfo{year}{2002}\natexlab{}.
\newblock \showarticletitle{Community structure in social and biological
  networks}.
\newblock \bibinfo{journal}{\emph{Proceedings of the National Academy of
  Sciences}} \bibinfo{volume}{99}, \bibinfo{number}{12} (\bibinfo{date}{June}
  \bibinfo{year}{2002}), \bibinfo{pages}{7821--7826}.
\newblock
\urldef\tempurl%
\url{https://doi.org/10.1073/pnas.122653799}
\showDOI{\tempurl}


\bibitem[\protect\citeauthoryear{Gratton, Sartenaer, and Toint}{Gratton
  et~al\mbox{.}}{2008}]%
        {gratton2008recursive}
\bibfield{author}{\bibinfo{person}{Serge Gratton}, \bibinfo{person}{Annick
  Sartenaer}, {and} \bibinfo{person}{Philippe~L Toint}.}
  \bibinfo{year}{2008}\natexlab{}.
\newblock \showarticletitle{Recursive trust-region methods for multiscale
  nonlinear optimization}.
\newblock \bibinfo{journal}{\emph{SIAM Journal on Optimization}}
  \bibinfo{volume}{19}, \bibinfo{number}{1} (\bibinfo{year}{2008}),
  \bibinfo{pages}{414--444}.
\newblock


\bibitem[\protect\citeauthoryear{Hagberg, Schult, and Swart}{Hagberg
  et~al\mbox{.}}{2008}]%
        {hagberg2008-rs}
\bibfield{author}{\bibinfo{person}{Aric~A. Hagberg}, \bibinfo{person}{Daniel~A.
  Schult}, {and} \bibinfo{person}{Pieter~J. Swart}.}
  \bibinfo{year}{2008}\natexlab{}.
\newblock \showarticletitle{Exploring Network Structure, Dynamics, and Function
  using {NetworkX}}. In \bibinfo{booktitle}{\emph{Proceedings of the 7th Python
  in Science Conference (SciPy 2008)}},
  \bibfield{editor}{\bibinfo{person}{Ga\"el Varoquaux}, \bibinfo{person}{Travis
  Vaught}, {and} \bibinfo{person}{Jarrod Millman}} (Eds.).
  \bibinfo{address}{Pasadena, CA USA}, \bibinfo{pages}{11--15}.
\newblock


\bibitem[\protect\citeauthoryear{Hager, Hungerford, and Safro}{Hager
  et~al\mbox{.}}{2018}]%
        {hager2018multilevel}
\bibfield{author}{\bibinfo{person}{William~W Hager}, \bibinfo{person}{James~T
  Hungerford}, {and} \bibinfo{person}{Ilya Safro}.}
  \bibinfo{year}{2018}\natexlab{}.
\newblock \showarticletitle{A multilevel bilinear programming algorithm for the
  vertex separator problem}.
\newblock \bibinfo{journal}{\emph{Computational Optimization and Applications}}
  \bibinfo{volume}{69}, \bibinfo{number}{1} (\bibinfo{year}{2018}),
  \bibinfo{pages}{189--223}.
\newblock


\bibitem[\protect\citeauthoryear{Hamerly, Inagaki, McMahon, Venturelli,
  Marandi, et~al\mbox{.}}{Hamerly et~al\mbox{.}}{2019}]%
        {Hamerly2019exp}
\bibfield{author}{\bibinfo{person}{Ryan Hamerly}, \bibinfo{person}{Takahiro
  Inagaki}, \bibinfo{person}{Peter~L. McMahon}, \bibinfo{person}{Davide
  Venturelli}, \bibinfo{person}{Alireza Marandi}, {et~al\mbox{.}}}
  \bibinfo{year}{2019}\natexlab{}.
\newblock \showarticletitle{Experimental investigation of performance
  differences between coherent Ising machines and a quantum annealer}.
\newblock \bibinfo{journal}{\emph{Science Advances}} \bibinfo{volume}{5},
  \bibinfo{number}{5} (\bibinfo{year}{2019}), \bibinfo{pages}{eaau0823}.
\newblock


\bibitem[\protect\citeauthoryear{Hart, Laird, Watson, Woodruff, Hackebeil,
  et~al\mbox{.}}{Hart et~al\mbox{.}}{2017}]%
        {hart2017pyomo}
\bibfield{author}{\bibinfo{person}{William~E Hart}, \bibinfo{person}{Carl~D
  Laird}, \bibinfo{person}{Jean-Paul Watson}, \bibinfo{person}{David~L
  Woodruff}, \bibinfo{person}{Gabriel~A Hackebeil}, {et~al\mbox{.}}}
  \bibinfo{year}{2017}\natexlab{}.
\newblock \bibinfo{booktitle}{\emph{Pyomo-optimization modeling in python}}.
  Vol.~\bibinfo{volume}{67}.
\newblock \bibinfo{publisher}{Springer}.
\newblock


\bibitem[\protect\citeauthoryear{Holtgrewe, Sanders, and Schulz}{Holtgrewe
  et~al\mbox{.}}{2010}]%
        {Holtgrewe2010}
\bibfield{author}{\bibinfo{person}{Manuel Holtgrewe}, \bibinfo{person}{Peter
  Sanders}, {and} \bibinfo{person}{Christian Schulz}.}
  \bibinfo{year}{2010}\natexlab{}.
\newblock \showarticletitle{Engineering a scalable high quality graph
  partitioner}. In \bibinfo{booktitle}{\emph{2010 {IEEE} International
  Symposium on Parallel {\&} Distributed Processing ({IPDPS})}}.
  \bibinfo{publisher}{{IEEE}}.
\newblock
\urldef\tempurl%
\url{https://doi.org/10.1109/ipdps.2010.5470485}
\showDOI{\tempurl}


\bibitem[\protect\citeauthoryear{Horst, Pardalos, and Van~Thoai}{Horst
  et~al\mbox{.}}{2000}]%
        {horst2000introduction}
\bibfield{author}{\bibinfo{person}{Reiner Horst}, \bibinfo{person}{Panos~M
  Pardalos}, {and} \bibinfo{person}{Nguyen Van~Thoai}.}
  \bibinfo{year}{2000}\natexlab{}.
\newblock \bibinfo{booktitle}{\emph{Introduction to global optimization}}.
\newblock \bibinfo{publisher}{Springer Science \& Business Media}.
\newblock


\bibitem[\protect\citeauthoryear{Huang, Szegedy, Zhang, Gao, Chen,
  et~al\mbox{.}}{Huang et~al\mbox{.}}{2019}]%
        {huang2019alibabacloud}
\bibfield{author}{\bibinfo{person}{Cupjin Huang}, \bibinfo{person}{Mario
  Szegedy}, \bibinfo{person}{Fang Zhang}, \bibinfo{person}{Xun Gao},
  \bibinfo{person}{Jianxin Chen}, {et~al\mbox{.}}}
  \bibinfo{year}{2019}\natexlab{}.
\newblock \bibinfo{title}{Alibaba Cloud Quantum Development Platform:
  Applications to Quantum Algorithm Design}.
\newblock
\newblock
\showeprint{arXiv:1909.02559}


\bibitem[\protect\citeauthoryear{John and Safro}{John and Safro}{2016}]%
        {safro:spars}
\bibfield{author}{\bibinfo{person}{Emmanuel John} {and} \bibinfo{person}{Ilya
  Safro}.} \bibinfo{year}{2016}\natexlab{}.
\newblock \showarticletitle{Single-and multi-level network sparsification by
  algebraic distance}.
\newblock \bibinfo{journal}{\emph{Journal of Complex Networks}}
  \bibinfo{volume}{5}, \bibinfo{number}{3} (\bibinfo{year}{2016}),
  \bibinfo{pages}{352--388}.
\newblock


\bibitem[\protect\citeauthoryear{Kandala, Mezzacapo, Temme, Takita, Brink,
  et~al\mbox{.}}{Kandala et~al\mbox{.}}{2017}]%
        {kandala2017hardware}
\bibfield{author}{\bibinfo{person}{Abhinav Kandala}, \bibinfo{person}{Antonio
  Mezzacapo}, \bibinfo{person}{Kristan Temme}, \bibinfo{person}{Maika Takita},
  \bibinfo{person}{Markus Brink}, {et~al\mbox{.}}}
  \bibinfo{year}{2017}\natexlab{}.
\newblock \showarticletitle{Hardware-efficient variational quantum eigensolver
  for small molecules and quantum magnets}.
\newblock \bibinfo{journal}{\emph{Nature}} \bibinfo{volume}{549},
  \bibinfo{number}{7671} (\bibinfo{year}{2017}), \bibinfo{pages}{242}.
\newblock


\bibitem[\protect\citeauthoryear{Karalekas, Tezak, Peterson, Ryan, da~Silva,
  et~al\mbox{.}}{Karalekas et~al\mbox{.}}{2020}]%
        {karalekas2020optimizedforvariational}
\bibfield{author}{\bibinfo{person}{Peter~J Karalekas},
  \bibinfo{person}{Nikolas~A Tezak}, \bibinfo{person}{Eric~C Peterson},
  \bibinfo{person}{Colm~A Ryan}, \bibinfo{person}{Marcus~P da Silva},
  {et~al\mbox{.}}} \bibinfo{year}{2020}\natexlab{}.
\newblock \showarticletitle{A quantum-classical cloud platform optimized for
  variational hybrid algorithms}.
\newblock \bibinfo{journal}{\emph{Quantum Science and Technology}}
  \bibinfo{volume}{5}, \bibinfo{number}{2} (\bibinfo{date}{April}
  \bibinfo{year}{2020}), \bibinfo{pages}{024003}.
\newblock
\urldef\tempurl%
\url{https://doi.org/10.1088/2058-9565/ab7559}
\showDOI{\tempurl}


\bibitem[\protect\citeauthoryear{Karypis and Kumar}{Karypis and Kumar}{1995}]%
        {KarypisKumar95b}
\bibfield{author}{\bibinfo{person}{G. Karypis} {and} \bibinfo{person}{V.
  Kumar}.} \bibinfo{year}{1995}\natexlab{}.
\newblock \bibinfo{booktitle}{\emph{Analysis of multilevel graph
  partitioning}}.
\newblock \bibinfo{type}{{T}echnical {R}eport} TR-95-037.
  \bibinfo{institution}{Computer Science Dept., Univ. of Minnesota},
  \bibinfo{address}{Minneapolis, MN}.
\newblock


\bibitem[\protect\citeauthoryear{Karypis and Kumar}{Karypis and Kumar}{1999}]%
        {KarypisKumar99fast}
\bibfield{author}{\bibinfo{person}{G. Karypis} {and} \bibinfo{person}{V.
  Kumar}.} \bibinfo{year}{1999}\natexlab{}.
\newblock \showarticletitle{A Fast and High Quality Multilevel Scheme for
  Partitioning Irregular Graphs}.
\newblock \bibinfo{journal}{\emph{SIAM Journal on Scientific Computing}}
  \bibinfo{volume}{20}, \bibinfo{number}{1} (\bibinfo{year}{1999}).
\newblock


\bibitem[\protect\citeauthoryear{Kato}{Kato}{1950}]%
        {Kato1950}
\bibfield{author}{\bibinfo{person}{Tosio Kato}.}
  \bibinfo{year}{1950}\natexlab{}.
\newblock \showarticletitle{On the Adiabatic Theorem of Quantum Mechanics}.
\newblock \bibinfo{journal}{\emph{Journal of the Physical Society of Japan}}
  \bibinfo{volume}{5}, \bibinfo{number}{6} (\bibinfo{date}{Nov.}
  \bibinfo{year}{1950}), \bibinfo{pages}{435--439}.
\newblock
\urldef\tempurl%
\url{https://doi.org/10.1143/jpsj.5.435}
\showDOI{\tempurl}


\bibitem[\protect\citeauthoryear{Katzgraber, Hamze, Zhu, Ochoa, and
  Munoz-Bauza}{Katzgraber et~al\mbox{.}}{2015}]%
        {Katzgraber2015}
\bibfield{author}{\bibinfo{person}{Helmut~G. Katzgraber},
  \bibinfo{person}{Firas Hamze}, \bibinfo{person}{Zheng Zhu},
  \bibinfo{person}{Andrew~J. Ochoa}, {and} \bibinfo{person}{H. Munoz-Bauza}.}
  \bibinfo{year}{2015}\natexlab{}.
\newblock \showarticletitle{Seeking Quantum Speedup Through Spin Glasses: The
  Good, the Bad, and the Ugly}.
\newblock \bibinfo{journal}{\emph{Physical Review X}} \bibinfo{volume}{5},
  \bibinfo{number}{3} (\bibinfo{date}{Sept.} \bibinfo{year}{2015}).
\newblock
\urldef\tempurl%
\url{https://doi.org/10.1103/physrevx.5.031026}
\showDOI{\tempurl}


\bibitem[\protect\citeauthoryear{Kelley}{Kelley}{1999}]%
        {kelley1999iterative}
\bibfield{author}{\bibinfo{person}{Carl~T Kelley}.}
  \bibinfo{year}{1999}\natexlab{}.
\newblock \bibinfo{booktitle}{\emph{Iterative methods for optimization}}.
\newblock \bibinfo{publisher}{SIAM}.
\newblock


\bibitem[\protect\citeauthoryear{Khairy, Shaydulin, Cincio, Alexeev, and
  Balaprakash}{Khairy et~al\mbox{.}}{2019}]%
        {khairy2019learning}
\bibfield{author}{\bibinfo{person}{Sami Khairy}, \bibinfo{person}{Ruslan
  Shaydulin}, \bibinfo{person}{Lukasz Cincio}, \bibinfo{person}{Yuri Alexeev},
  {and} \bibinfo{person}{Prasanna Balaprakash}.}
  \bibinfo{year}{2019}\natexlab{}.
\newblock \showarticletitle{Learning to Optimize Variational Quantum Circuits
  to Solve Combinatorial Problems}.
\newblock \bibinfo{journal}{\emph{Proceedings of the Thirty-Forth AAAI
  Conference on Artificial Intelligence (AAAI-20)}} (\bibinfo{year}{2019}).
\newblock


\bibitem[\protect\citeauthoryear{King, Raymond, Lanting, Isakov, Mohseni,
  et~al\mbox{.}}{King et~al\mbox{.}}{2019a}]%
        {King2019scale}
\bibfield{author}{\bibinfo{person}{Andrew~D King}, \bibinfo{person}{Jack
  Raymond}, \bibinfo{person}{Trevor Lanting}, \bibinfo{person}{Sergei~V
  Isakov}, \bibinfo{person}{Masoud Mohseni}, {et~al\mbox{.}}}
  \bibinfo{year}{2019}\natexlab{a}.
\newblock \showarticletitle{Scaling advantage in quantum simulation of
  geometrically frustrated magnets}.
\newblock \bibinfo{journal}{\emph{arXiv preprint arXiv:1911.03446}}
  (\bibinfo{year}{2019}).
\newblock


\bibitem[\protect\citeauthoryear{King, Yarkoni, Raymond, Ozfidan, King,
  et~al\mbox{.}}{King et~al\mbox{.}}{2019b}]%
        {King2019}
\bibfield{author}{\bibinfo{person}{James King}, \bibinfo{person}{Sheir
  Yarkoni}, \bibinfo{person}{Jack Raymond}, \bibinfo{person}{Isil Ozfidan},
  \bibinfo{person}{Andrew~D. King}, {et~al\mbox{.}}}
  \bibinfo{year}{2019}\natexlab{b}.
\newblock \showarticletitle{Quantum Annealing amid Local Ruggedness and Global
  Frustration}.
\newblock \bibinfo{journal}{\emph{Journal of the Physical Society of Japan}}
  \bibinfo{volume}{88}, \bibinfo{number}{6} (\bibinfo{date}{June}
  \bibinfo{year}{2019}), \bibinfo{pages}{061007}.
\newblock
\urldef\tempurl%
\url{https://doi.org/10.7566/jpsj.88.061007}
\showDOI{\tempurl}


\bibitem[\protect\citeauthoryear{Kunegis}{Kunegis}{2013a}]%
        {KONECT13}
\bibfield{author}{\bibinfo{person}{J{\'e}r{\^o}me Kunegis}.}
  \bibinfo{year}{2013}\natexlab{a}.
\newblock \showarticletitle{{KONECT} -- The Koblenz Network Collection}. In
  \bibinfo{booktitle}{\emph{Web Observatory Workshop}}.
  \bibinfo{pages}{1343--1350}.
\newblock


\bibitem[\protect\citeauthoryear{Kunegis}{Kunegis}{2013b}]%
        {kunegis2013konect_rs}
\bibfield{author}{\bibinfo{person}{J{\'e}r{\^o}me Kunegis}.}
  \bibinfo{year}{2013}\natexlab{b}.
\newblock \showarticletitle{Konect: the koblenz network collection}. In
  \bibinfo{booktitle}{\emph{Proceedings of the 22nd International Conference on
  World Wide Web}}. ACM, \bibinfo{pages}{1343--1350}.
\newblock


\bibitem[\protect\citeauthoryear{Lancichinetti, Fortunato, and
  Radicchi}{Lancichinetti et~al\mbox{.}}{2008}]%
        {Lancichinetti2008}
\bibfield{author}{\bibinfo{person}{Andrea Lancichinetti},
  \bibinfo{person}{Santo Fortunato}, {and} \bibinfo{person}{Filippo Radicchi}.}
  \bibinfo{year}{2008}\natexlab{}.
\newblock \showarticletitle{Benchmark graphs for testing community detection
  algorithms}.
\newblock \bibinfo{journal}{\emph{Physical Review E}} \bibinfo{volume}{78},
  \bibinfo{number}{4} (\bibinfo{date}{Oct.} \bibinfo{year}{2008}).
\newblock
\urldef\tempurl%
\url{https://doi.org/10.1103/physreve.78.046110}
\showDOI{\tempurl}


\bibitem[\protect\citeauthoryear{Leyffer and Safro}{Leyffer and Safro}{2013}]%
        {leyffer2013fast}
\bibfield{author}{\bibinfo{person}{Sven Leyffer} {and} \bibinfo{person}{Ilya
  Safro}.} \bibinfo{year}{2013}\natexlab{}.
\newblock \showarticletitle{Fast response to infection spread and cyber attacks
  on large-scale networks}.
\newblock \bibinfo{journal}{\emph{Journal of Complex Networks}}
  \bibinfo{volume}{1}, \bibinfo{number}{2} (\bibinfo{year}{2013}),
  \bibinfo{pages}{183--199}.
\newblock


\bibitem[\protect\citeauthoryear{Livne and Brandt}{Livne and Brandt}{2012}]%
        {livne2012lean}
\bibfield{author}{\bibinfo{person}{Oren~E Livne} {and} \bibinfo{person}{Achi
  Brandt}.} \bibinfo{year}{2012}\natexlab{}.
\newblock \showarticletitle{Lean algebraic multigrid (LAMG): Fast graph
  Laplacian linear solver}.
\newblock \bibinfo{journal}{\emph{SIAM Journal on Scientific Computing}}
  \bibinfo{volume}{34}, \bibinfo{number}{4} (\bibinfo{year}{2012}),
  \bibinfo{pages}{B499--B522}.
\newblock


\bibitem[\protect\citeauthoryear{McClean, Boixo, Smelyanskiy, Babbush, and
  Neven}{McClean et~al\mbox{.}}{2018}]%
        {McClean2018barrendoi}
\bibfield{author}{\bibinfo{person}{Jarrod~R. McClean}, \bibinfo{person}{Sergio
  Boixo}, \bibinfo{person}{Vadim~N. Smelyanskiy}, \bibinfo{person}{Ryan
  Babbush}, {and} \bibinfo{person}{Hartmut Neven}.}
  \bibinfo{year}{2018}\natexlab{}.
\newblock \showarticletitle{Barren plateaus in quantum neural network training
  landscapes}.
\newblock \bibinfo{journal}{\emph{Nature Communications}} \bibinfo{volume}{9},
  \bibinfo{number}{1} (\bibinfo{date}{Nov.} \bibinfo{year}{2018}).
\newblock
\urldef\tempurl%
\url{https://doi.org/10.1038/s41467-018-07090-4}
\showDOI{\tempurl}


\bibitem[\protect\citeauthoryear{McGeoch}{McGeoch}{2019}]%
        {mcgeoch2019performance}
\bibfield{author}{\bibinfo{person}{Catherine~C McGeoch}.}
  \bibinfo{year}{2019}\natexlab{}.
\newblock \showarticletitle{Principles and Guidelines for Quantum Performance
  Analysis}.
\newblock \bibinfo{journal}{\emph{Quantum Technology and Optimization Problems.
  (QTOP 2019). Lecture Notes in Computer Science}}  \bibinfo{volume}{411413}
  (\bibinfo{year}{2019}), \bibinfo{pages}{36--47}.
\newblock


\bibitem[\protect\citeauthoryear{Migdalas, Pardalos, and V{\"a}rbrand}{Migdalas
  et~al\mbox{.}}{2013}]%
        {migdalas2013multilevel}
\bibfield{author}{\bibinfo{person}{Athanasios Migdalas},
  \bibinfo{person}{Panos~M Pardalos}, {and} \bibinfo{person}{Peter
  V{\"a}rbrand}.} \bibinfo{year}{2013}\natexlab{}.
\newblock \bibinfo{booktitle}{\emph{Multilevel optimization: algorithms and
  applications}}. Vol.~\bibinfo{volume}{20}.
\newblock \bibinfo{publisher}{Springer Science \& Business Media}.
\newblock


\bibitem[\protect\citeauthoryear{Murty and Kabadi}{Murty and Kabadi}{1987}]%
        {murty1987}
\bibfield{author}{\bibinfo{person}{K.~G. Murty} {and} \bibinfo{person}{S.~N.
  Kabadi}.} \bibinfo{year}{1987}\natexlab{}.
\newblock \showarticletitle{Some {NP}-complete problems in quadratic and linear
  programming}.
\newblock \bibinfo{journal}{\emph{Math. Program.}}  \bibinfo{volume}{39}
  (\bibinfo{year}{1987}), \bibinfo{pages}{117--129}.
\newblock


\bibitem[\protect\citeauthoryear{Nannicini}{Nannicini}{2019}]%
        {Nannicini2019performance}
\bibfield{author}{\bibinfo{person}{Giacomo Nannicini}.}
  \bibinfo{year}{2019}\natexlab{}.
\newblock \showarticletitle{Performance of hybrid quantum-classical variational
  heuristics for combinatorial optimization}.
\newblock \bibinfo{journal}{\emph{Physical Review E}} \bibinfo{volume}{99},
  \bibinfo{number}{1} (\bibinfo{date}{Jan.} \bibinfo{year}{2019}).
\newblock
\urldef\tempurl%
\url{https://doi.org/10.1103/physreve.99.013304}
\showDOI{\tempurl}


\bibitem[\protect\citeauthoryear{Napp, Placa, Dalzell, Brandao, and
  Harrow}{Napp et~al\mbox{.}}{2019}]%
        {napp2019efficientshallow}
\bibfield{author}{\bibinfo{person}{John Napp}, \bibinfo{person}{Rolando L.~La
  Placa}, \bibinfo{person}{Alexander~M. Dalzell}, \bibinfo{person}{Fernando G.
  S.~L. Brandao}, {and} \bibinfo{person}{Aram~W. Harrow}.}
  \bibinfo{year}{2019}\natexlab{}.
\newblock \bibinfo{title}{Efficient classical simulation of random shallow 2D
  quantum circuits}.
\newblock
\newblock
\showeprint{arXiv:2001.00021}


\bibitem[\protect\citeauthoryear{Negre, Ushijima-Mwesigwa, and
  Mniszewski}{Negre et~al\mbox{.}}{2020}]%
        {negre2019detecting}
\bibfield{author}{\bibinfo{person}{Christian F.~A. Negre},
  \bibinfo{person}{Hayato Ushijima-Mwesigwa}, {and} \bibinfo{person}{Susan~M.
  Mniszewski}.} \bibinfo{year}{2020}\natexlab{}.
\newblock \showarticletitle{Detecting multiple communities using quantum
  annealing on the D-Wave system}.
\newblock \bibinfo{journal}{\emph{PLOS ONE}}  \bibinfo{volume}{15}
  (\bibinfo{date}{02} \bibinfo{year}{2020}), \bibinfo{pages}{1--14}.
\newblock
\urldef\tempurl%
\url{https://doi.org/10.1371/journal.pone.0227538}
\showURL{%
\tempurl}


\bibitem[\protect\citeauthoryear{Newman}{Newman}{2006}]%
        {Newman06modularity}
\bibfield{author}{\bibinfo{person}{M.~E.~J. Newman}.}
  \bibinfo{year}{2006}\natexlab{}.
\newblock \showarticletitle{Modularity and community structure in networks}.
\newblock \bibinfo{journal}{\emph{Proceedings of National Academy of Sciences}}
   \bibinfo{volume}{103} (\bibinfo{year}{2006}), \bibinfo{pages}{8577}.
\newblock
\urldef\tempurl%
\url{doi:10.1073/pnas.0601602103}
\showURL{%
\tempurl}


\bibitem[\protect\citeauthoryear{Novikov, Hinkey, Disseler, Basham, Albash,
  et~al\mbox{.}}{Novikov et~al\mbox{.}}{2018}]%
        {novikov2018exploring}
\bibfield{author}{\bibinfo{person}{Sergey Novikov}, \bibinfo{person}{Robert
  Hinkey}, \bibinfo{person}{Steven Disseler}, \bibinfo{person}{James~I Basham},
  \bibinfo{person}{Tameem Albash}, {et~al\mbox{.}}}
  \bibinfo{year}{2018}\natexlab{}.
\newblock \showarticletitle{Exploring More-Coherent Quantum Annealing}.
\newblock \bibinfo{journal}{\emph{arXiv preprint arXiv:1809.04485}}
  (\bibinfo{year}{2018}).
\newblock


\bibitem[\protect\citeauthoryear{Optimization}{Optimization}{2014}]%
        {optimization2014inc-rs}
\bibfield{author}{\bibinfo{person}{Gurobi Optimization}.}
  \bibinfo{year}{2014}\natexlab{}.
\newblock \showarticletitle{Inc.,“Gurobi optimizer reference manual,”
  2015}.
\newblock \bibinfo{howpublished}{\url{http://www. gurobi. com}}.
\newblock  (\bibinfo{year}{2014}).
\newblock


\bibitem[\protect\citeauthoryear{Otterbach, Manenti, Alidoust, Bestwick, Block,
  et~al\mbox{.}}{Otterbach et~al\mbox{.}}{2017}]%
        {otterbach2017unsupervised}
\bibfield{author}{\bibinfo{person}{JS Otterbach}, \bibinfo{person}{R Manenti},
  \bibinfo{person}{N Alidoust}, \bibinfo{person}{A Bestwick},
  \bibinfo{person}{M Block}, {et~al\mbox{.}}} \bibinfo{year}{2017}\natexlab{}.
\newblock \showarticletitle{Unsupervised Machine Learning on a Hybrid Quantum
  Computer}.
\newblock \bibinfo{journal}{\emph{arXiv preprint arXiv:1712.05771}}
  (\bibinfo{year}{2017}).
\newblock


\bibitem[\protect\citeauthoryear{Pagano, Bapat, Becker, Collins, De,
  et~al\mbox{.}}{Pagano et~al\mbox{.}}{2019}]%
        {pagano2019quantum}
\bibfield{author}{\bibinfo{person}{G Pagano}, \bibinfo{person}{A Bapat},
  \bibinfo{person}{P Becker}, \bibinfo{person}{KS Collins}, \bibinfo{person}{A
  De}, {et~al\mbox{.}}} \bibinfo{year}{2019}\natexlab{}.
\newblock \showarticletitle{Quantum Approximate Optimization with a Trapped-Ion
  Quantum Simulator}.
\newblock \bibinfo{journal}{\emph{arXiv preprint arXiv:1906.02700}}
  (\bibinfo{year}{2019}).
\newblock


\bibitem[\protect\citeauthoryear{Par{\'e}s, Gasulla, Vilalta, Moreno,
  Ayguad{\'e}, et~al\mbox{.}}{Par{\'e}s et~al\mbox{.}}{2017}]%
        {pares2017fluid}
\bibfield{author}{\bibinfo{person}{Ferran Par{\'e}s},
  \bibinfo{person}{Dario~Garcia Gasulla}, \bibinfo{person}{Armand Vilalta},
  \bibinfo{person}{Jonatan Moreno}, \bibinfo{person}{Eduard Ayguad{\'e}},
  {et~al\mbox{.}}} \bibinfo{year}{2017}\natexlab{}.
\newblock \showarticletitle{Fluid communities: a competitive, scalable and
  diverse community detection algorithm}. In
  \bibinfo{booktitle}{\emph{International Conference on Complex Networks and
  their Applications}}. Springer, \bibinfo{pages}{229--240}.
\newblock


\bibitem[\protect\citeauthoryear{Pedregosa, Varoquaux, Gramfort, Michel,
  Thirion, et~al\mbox{.}}{Pedregosa et~al\mbox{.}}{2011}]%
        {scikit-learn}
\bibfield{author}{\bibinfo{person}{F. Pedregosa}, \bibinfo{person}{G.
  Varoquaux}, \bibinfo{person}{A. Gramfort}, \bibinfo{person}{V. Michel},
  \bibinfo{person}{B. Thirion}, {et~al\mbox{.}}}
  \bibinfo{year}{2011}\natexlab{}.
\newblock \showarticletitle{Scikit-learn: Machine Learning in {P}ython}.
\newblock \bibinfo{journal}{\emph{Journal of Machine Learning Research}}
  \bibinfo{volume}{12} (\bibinfo{year}{2011}), \bibinfo{pages}{2825--2830}.
\newblock


\bibitem[\protect\citeauthoryear{Peruzzo, McClean, Shadbolt, Yung, Zhou,
  et~al\mbox{.}}{Peruzzo et~al\mbox{.}}{2014}]%
        {peruzzo2014variational}
\bibfield{author}{\bibinfo{person}{Alberto Peruzzo}, \bibinfo{person}{Jarrod
  McClean}, \bibinfo{person}{Peter Shadbolt}, \bibinfo{person}{Man-Hong Yung},
  \bibinfo{person}{Xiao-Qi Zhou}, {et~al\mbox{.}}}
  \bibinfo{year}{2014}\natexlab{}.
\newblock \showarticletitle{A variational eigenvalue solver on a photonic
  quantum processor}.
\newblock \bibinfo{journal}{\emph{Nature communications}}  \bibinfo{volume}{5}
  (\bibinfo{year}{2014}), \bibinfo{pages}{4213}.
\newblock


\bibitem[\protect\citeauthoryear{Pichler, Wang, Zhou, Choi, and Lukin}{Pichler
  et~al\mbox{.}}{2018}]%
        {pichler2018quantum}
\bibfield{author}{\bibinfo{person}{Hannes Pichler}, \bibinfo{person}{Sheng-Tao
  Wang}, \bibinfo{person}{Leo Zhou}, \bibinfo{person}{Soonwon Choi}, {and}
  \bibinfo{person}{Mikhail~D Lukin}.} \bibinfo{year}{2018}\natexlab{}.
\newblock \showarticletitle{Quantum Optimization for Maximum Independent Set
  Using Rydberg Atom Arrays}.
\newblock \bibinfo{journal}{\emph{arXiv preprint arXiv:1808.10816}}
  (\bibinfo{year}{2018}).
\newblock


\bibitem[\protect\citeauthoryear{Ron, Safro, and Brandt}{Ron
  et~al\mbox{.}}{2010}]%
        {Ron2010}
\bibfield{author}{\bibinfo{person}{Dorit Ron}, \bibinfo{person}{Ilya Safro},
  {and} \bibinfo{person}{Achi Brandt}.} \bibinfo{year}{2010}\natexlab{}.
\newblock \showarticletitle{A Fast Multigrid Algorithm for Energy Minimization
  under Planar Density Constraints}.
\newblock \bibinfo{journal}{\emph{{SIAM} {M}ultiscale {M}odeling {\&}
  {S}imulation}} \bibinfo{volume}{8}, \bibinfo{number}{5}
  (\bibinfo{year}{2010}), \bibinfo{pages}{1599--1620}.
\newblock


\bibitem[\protect\citeauthoryear{Ron, Safro, and Brandt}{Ron
  et~al\mbox{.}}{2011}]%
        {safro:relaxml}
\bibfield{author}{\bibinfo{person}{Dorit Ron}, \bibinfo{person}{Ilya Safro},
  {and} \bibinfo{person}{Achi Brandt}.} \bibinfo{year}{2011}\natexlab{}.
\newblock \showarticletitle{Relaxation-based coarsening and multiscale graph
  organization}.
\newblock \bibinfo{journal}{\emph{Multiscale Modeling \& Simulation}}
  \bibinfo{volume}{9}, \bibinfo{number}{1} (\bibinfo{year}{2011}),
  \bibinfo{pages}{407--423}.
\newblock


\bibitem[\protect\citeauthoryear{Rowan}{Rowan}{1990}]%
        {rowan1991functional}
\bibfield{author}{\bibinfo{person}{Thomas~Harvey Rowan}.}
  \bibinfo{year}{1990}\natexlab{}.
\newblock \emph{\bibinfo{title}{Functional stability analysis of numerical
  algorithms.}}
\newblock \bibinfo{thesistype}{Ph.D. Dissertation}.
  \bibinfo{school}{{U}niversity of {T}exas at {A}ustin}.
\newblock


\bibitem[\protect\citeauthoryear{Sadrfaridpour, Razzaghi, and
  Safro}{Sadrfaridpour et~al\mbox{.}}{2019}]%
        {sadrfaridpour2019engineering}
\bibfield{author}{\bibinfo{person}{Ehsan Sadrfaridpour},
  \bibinfo{person}{Talayeh Razzaghi}, {and} \bibinfo{person}{Ilya Safro}.}
  \bibinfo{year}{2019}\natexlab{}.
\newblock \showarticletitle{Engineering fast multilevel support vector
  machines}.
\newblock \bibinfo{journal}{\emph{Machine Learning}} (\bibinfo{year}{2019}),
  \bibinfo{pages}{1--39}.
\newblock


\bibitem[\protect\citeauthoryear{Sadrfaridpour, Sandeep, Kennedy, Luckow,
  Razzaghi, et~al\mbox{.}}{Sadrfaridpour et~al\mbox{.}}{2017}]%
        {sadrfaridpour_ehsan2017}
\bibfield{author}{\bibinfo{person}{Ehsan Sadrfaridpour},
  \bibinfo{person}{Jeereddy Sandeep}, \bibinfo{person}{Ken Kennedy},
  \bibinfo{person}{Andre Luckow}, \bibinfo{person}{Talayeh Razzaghi},
  {et~al\mbox{.}}} \bibinfo{year}{2017}\natexlab{}.
\newblock \showarticletitle{Algebraic multigrid support vector machines}. In
  \bibinfo{booktitle}{\emph{ESANN 2017 proceedings, European Symposium on
  Artificial Neural Networks, Computational Intelligence and Machine
  Learning}}. \bibinfo{publisher}{ESANN}, \bibinfo{address}{Bruges, Belgium}.
\newblock
\showISBNx{9782875870391}
\urldef\tempurl%
\url{https://www.elen.ucl.ac.be/Proceedings/esann/esannpdf/es2017-37.pdf}
\showURL{%
\tempurl}


\bibitem[\protect\citeauthoryear{Safro, Ron, and Brandt}{Safro
  et~al\mbox{.}}{2006}]%
        {Safro2006}
\bibfield{author}{\bibinfo{person}{Ilya Safro}, \bibinfo{person}{Dorit Ron},
  {and} \bibinfo{person}{Achi Brandt}.} \bibinfo{year}{2006}\natexlab{}.
\newblock \showarticletitle{Graph minimum linear arrangement by multilevel
  weighted edge contractions}.
\newblock \bibinfo{journal}{\emph{J. Algorithms}} \bibinfo{volume}{60},
  \bibinfo{number}{1} (\bibinfo{year}{2006}), \bibinfo{pages}{24--41}.
\newblock


\bibitem[\protect\citeauthoryear{Safro, Ron, and Brandt}{Safro
  et~al\mbox{.}}{2008}]%
        {SafroRB08}
\bibfield{author}{\bibinfo{person}{Ilya Safro}, \bibinfo{person}{Dorit Ron},
  {and} \bibinfo{person}{Achi Brandt}.} \bibinfo{year}{2008}\natexlab{}.
\newblock \showarticletitle{Multilevel algorithms for linear ordering
  problems}.
\newblock \bibinfo{journal}{\emph{ACM Journal of Experimental Algorithmics}}
  \bibinfo{volume}{13} (\bibinfo{year}{2008}).
\newblock


\bibitem[\protect\citeauthoryear{Safro, Sanders, and Schulz}{Safro
  et~al\mbox{.}}{2015}]%
        {amg-sss12}
\bibfield{author}{\bibinfo{person}{Ilya Safro}, \bibinfo{person}{Peter
  Sanders}, {and} \bibinfo{person}{Christian Schulz}.}
  \bibinfo{year}{2015}\natexlab{}.
\newblock \showarticletitle{Advanced coarsening schemes for graph
  partitioning}.
\newblock \bibinfo{journal}{\emph{ACM Journal of Experimental Algorithmics
  (JEA)}}  \bibinfo{volume}{19} (\bibinfo{year}{2015}), \bibinfo{pages}{2--2}.
\newblock


\bibitem[\protect\citeauthoryear{Safro and Temkin}{Safro and Temkin}{2011}]%
        {Safro2011}
\bibfield{author}{\bibinfo{person}{Ilya Safro} {and} \bibinfo{person}{Boris
  Temkin}.} \bibinfo{year}{2011}\natexlab{}.
\newblock \showarticletitle{Multiscale approach for the network
  compression-friendly ordering}.
\newblock \bibinfo{journal}{\emph{J. Discrete Algorithms}} \bibinfo{volume}{9},
  \bibinfo{number}{2} (\bibinfo{year}{2011}), \bibinfo{pages}{190--202}.
\newblock


\bibitem[\protect\citeauthoryear{Sanders and Schulz}{Sanders and
  Schulz}{2013}]%
        {sandersschulz2013}
\bibfield{author}{\bibinfo{person}{Peter Sanders} {and}
  \bibinfo{person}{Christian Schulz}.} \bibinfo{year}{2013}\natexlab{}.
\newblock \showarticletitle{{Think Locally, Act Globally: Highly Balanced Graph
  Partitioning}}. In \bibinfo{booktitle}{\emph{Proceedings of the 12th
  International Symposium on Experimental Algorithms (SEA'13)}}
  \emph{(\bibinfo{series}{LNCS})}, Vol.~\bibinfo{volume}{7933}.
  \bibinfo{publisher}{Springer}, \bibinfo{pages}{164--175}.
\newblock


\bibitem[\protect\citeauthoryear{Sanders, Schulz, and Wagner}{Sanders
  et~al\mbox{.}}{2014}]%
        {sanders2014benchmarking}
\bibfield{author}{\bibinfo{person}{Peter Sanders}, \bibinfo{person}{Christian
  Schulz}, {and} \bibinfo{person}{Dorothea Wagner}.}
  \bibinfo{year}{2014}\natexlab{}.
\newblock \showarticletitle{Benchmarking for graph clustering and
  partitioning}.
\newblock  (\bibinfo{year}{2014}).
\newblock


\bibitem[\protect\citeauthoryear{Shaydulin and Alexeev}{Shaydulin and
  Alexeev}{2019}]%
        {Shaydulin2019EvaluatingDOI}
\bibfield{author}{\bibinfo{person}{Ruslan Shaydulin} {and}
  \bibinfo{person}{Yuri Alexeev}.} \bibinfo{year}{2019}\natexlab{}.
\newblock \showarticletitle{Evaluating Quantum Approximate Optimization
  Algorithm: A Case Study}. In \bibinfo{booktitle}{\emph{2019 Tenth
  International Green and Sustainable Computing Conference ({IGSC})}}.
  \bibinfo{publisher}{{IEEE}}.
\newblock
\urldef\tempurl%
\url{https://doi.org/10.1109/igsc48788.2019.8957201}
\showDOI{\tempurl}


\bibitem[\protect\citeauthoryear{Shaydulin, Chen, and Safro}{Shaydulin
  et~al\mbox{.}}{2019a}]%
        {shaydulin2019algdist}
\bibfield{author}{\bibinfo{person}{Ruslan Shaydulin}, \bibinfo{person}{Jie
  Chen}, {and} \bibinfo{person}{Ilya Safro}.} \bibinfo{year}{2019}\natexlab{a}.
\newblock \showarticletitle{Relaxation-Based Coarsening for Multilevel
  Hypergraph Partitioning}.
\newblock \bibinfo{journal}{\emph{{SIAM Multiscale Modeling and Simulation}}}
  \bibinfo{volume}{17} (\bibinfo{year}{2019}), \bibinfo{pages}{482--506}.
\newblock
Issue 1.


\bibitem[\protect\citeauthoryear{Shaydulin and Safro}{Shaydulin and
  Safro}{2018}]%
        {shaydulin_et_al:LIPIcs:2018:8937}
\bibfield{author}{\bibinfo{person}{Ruslan Shaydulin} {and}
  \bibinfo{person}{Ilya Safro}.} \bibinfo{year}{2018}\natexlab{}.
\newblock \showarticletitle{{Aggregative Coarsening for Multilevel Hypergraph
  Partitioning}}. In \bibinfo{booktitle}{\emph{17th International Symposium on
  Experimental Algorithms (SEA 2018)}} \emph{(\bibinfo{series}{Leibniz
  International Proceedings in Informatics (LIPIcs)})},
  \bibfield{editor}{\bibinfo{person}{Gianlorenzo D'Angelo}} (Ed.),
  Vol.~\bibinfo{volume}{103}. \bibinfo{publisher}{Schloss
  Dagstuhl--Leibniz-Zentrum fuer Informatik}, \bibinfo{address}{Dagstuhl,
  Germany}, \bibinfo{pages}{2:1--2:15}.
\newblock
\showISBNx{978-3-95977-070-5}
\showISSN{1868-8969}
\urldef\tempurl%
\url{https://doi.org/10.4230/LIPIcs.SEA.2018.2}
\showDOI{\tempurl}


\bibitem[\protect\citeauthoryear{Shaydulin, Safro, and Larson}{Shaydulin
  et~al\mbox{.}}{2019b}]%
        {Shaydulin2019MultistartDOI}
\bibfield{author}{\bibinfo{person}{Ruslan Shaydulin}, \bibinfo{person}{Ilya
  Safro}, {and} \bibinfo{person}{Jeffrey Larson}.}
  \bibinfo{year}{2019}\natexlab{b}.
\newblock \showarticletitle{Multistart Methods for Quantum Approximate
  optimization}. In \bibinfo{booktitle}{\emph{2019 {IEEE} High Performance
  Extreme Computing Conference ({HPEC})}}. \bibinfo{publisher}{{IEEE}}.
\newblock
\urldef\tempurl%
\url{https://doi.org/10.1109/hpec.2019.8916288}
\showDOI{\tempurl}


\bibitem[\protect\citeauthoryear{Shaydulin, Ushijima-Mwesigwa, Negre, Safro,
  Mniszewski, et~al\mbox{.}}{Shaydulin et~al\mbox{.}}{2019c}]%
        {Shaydulin2019}
\bibfield{author}{\bibinfo{person}{Ruslan Shaydulin}, \bibinfo{person}{Hayato
  Ushijima-Mwesigwa}, \bibinfo{person}{Christian F.~A. Negre},
  \bibinfo{person}{Ilya Safro}, \bibinfo{person}{Susan~M. Mniszewski},
  {et~al\mbox{.}}} \bibinfo{year}{2019}\natexlab{c}.
\newblock \showarticletitle{A Hybrid Approach for Solving Optimization Problems
  on Small Quantum Computers}.
\newblock \bibinfo{journal}{\emph{Computer}} \bibinfo{volume}{52},
  \bibinfo{number}{6} (\bibinfo{date}{June} \bibinfo{year}{2019}),
  \bibinfo{pages}{18--26}.
\newblock
\urldef\tempurl%
\url{https://doi.org/10.1109/mc.2019.2908942}
\showDOI{\tempurl}


\bibitem[\protect\citeauthoryear{Shaydulin, Ushijima-Mwesigwa, Safro,
  Mniszewski, and Alexeev}{Shaydulin et~al\mbox{.}}{2018}]%
        {shaydulin2018community}
\bibfield{author}{\bibinfo{person}{Ruslan Shaydulin}, \bibinfo{person}{Hayato
  Ushijima-Mwesigwa}, \bibinfo{person}{Ilya Safro}, \bibinfo{person}{Susan
  Mniszewski}, {and} \bibinfo{person}{Yuri Alexeev}.}
  \bibinfo{year}{2018}\natexlab{}.
\newblock \showarticletitle{Community Detection Across Emerging Quantum
  Architectures}.
\newblock \bibinfo{journal}{\emph{Proceedings of the 3rd International Workshop
  on Post Moore's Era Supercomputing}} (\bibinfo{year}{2018}).
\newblock


\bibitem[\protect\citeauthoryear{Shaydulin, Ushijima-Mwesigwa, Safro,
  Mniszewski, and Alexeev}{Shaydulin et~al\mbox{.}}{2019d}]%
        {shaydulin2018network}
\bibfield{author}{\bibinfo{person}{Ruslan Shaydulin}, \bibinfo{person}{Hayato
  Ushijima-Mwesigwa}, \bibinfo{person}{Ilya Safro}, \bibinfo{person}{Susan
  Mniszewski}, {and} \bibinfo{person}{Yuri Alexeev}.}
  \bibinfo{year}{2019}\natexlab{d}.
\newblock \showarticletitle{Network Community Detection On Small Quantum
  Computers}.
\newblock \bibinfo{journal}{\emph{accepted in Advanced Quantum Technologies,
  preprint arXiv:1810.12484}} (\bibinfo{year}{2019}).
\newblock
\urldef\tempurl%
\url{https://doi.org/10.1002/qute.201900029}
\showURL{%
\tempurl}


\bibitem[\protect\citeauthoryear{Shi and Malik}{Shi and Malik}{2000}]%
        {shi2000normalized}
\bibfield{author}{\bibinfo{person}{Jianbo Shi} {and} \bibinfo{person}{Jitendra
  Malik}.} \bibinfo{year}{2000}\natexlab{}.
\newblock \showarticletitle{Normalized cuts and image segmentation}.
\newblock \bibinfo{journal}{\emph{Departmental Papers (CIS)}}
  (\bibinfo{year}{2000}), \bibinfo{pages}{107}.
\newblock


\bibitem[\protect\citeauthoryear{Soper, Walshaw, and Cross}{Soper
  et~al\mbox{.}}{2004}]%
        {Soper2004}
\bibfield{author}{\bibinfo{person}{A.J. Soper}, \bibinfo{person}{C. Walshaw},
  {and} \bibinfo{person}{M. Cross}.} \bibinfo{year}{2004}\natexlab{}.
\newblock \showarticletitle{A Combined Evolutionary Search and Multilevel
  Optimisation Approach to Graph-Partitioning}.
\newblock \bibinfo{journal}{\emph{Journal of Global Optimization}}
  \bibinfo{volume}{29}, \bibinfo{number}{2} (\bibinfo{date}{June}
  \bibinfo{year}{2004}), \bibinfo{pages}{225--241}.
\newblock
\urldef\tempurl%
\url{https://doi.org/10.1023/b:jogo.0000042115.44455.f3}
\showDOI{\tempurl}


\bibitem[\protect\citeauthoryear{Streif and Leib}{Streif and Leib}{2020}]%
        {Streif2020}
\bibfield{author}{\bibinfo{person}{Michael Streif} {and}
  \bibinfo{person}{Martin Leib}.} \bibinfo{year}{2020}\natexlab{}.
\newblock \showarticletitle{Training the quantum approximate optimization
  algorithm without access to a quantum processing unit}.
\newblock \bibinfo{journal}{\emph{Quantum Science and Technology}}
  \bibinfo{volume}{5}, \bibinfo{number}{3} (\bibinfo{date}{May}
  \bibinfo{year}{2020}), \bibinfo{pages}{034008}.
\newblock
\urldef\tempurl%
\url{https://doi.org/10.1088/2058-9565/ab8c2b}
\showDOI{\tempurl}


\bibitem[\protect\citeauthoryear{Szegedy}{Szegedy}{2019}]%
        {Szegedy2019qaoaenergies}
\bibfield{author}{\bibinfo{person}{Mario Szegedy}.}
  \bibinfo{year}{2019}\natexlab{}.
\newblock \bibinfo{title}{What do QAOA energies reveal about graphs?}
\newblock
\newblock
\showeprint{arXiv:1912.12277}


\bibitem[\protect\citeauthoryear{Tange}{Tange}{2018}]%
        {tange_ole_2018_1146014}
\bibfield{author}{\bibinfo{person}{Ole Tange}.}
  \bibinfo{year}{2018}\natexlab{}.
\newblock \bibinfo{booktitle}{\emph{GNU Parallel 2018}}.
\newblock \bibinfo{publisher}{Ole Tange}.
\newblock
\showISBNx{9781387509881}
\urldef\tempurl%
\url{https://doi.org/10.5281/zenodo.1146014}
\showDOI{\tempurl}


\bibitem[\protect\citeauthoryear{Terry, Akrobotu, Negre, and Mniszewski}{Terry
  et~al\mbox{.}}{2020}]%
        {Terry2020isomersearch}
\bibfield{author}{\bibinfo{person}{Jason~P Terry}, \bibinfo{person}{Prosper~D
  Akrobotu}, \bibinfo{person}{Christian F~A Negre}, {and}
  \bibinfo{person}{Susan~M Mniszewski}.} \bibinfo{year}{2020}\natexlab{}.
\newblock \showarticletitle{Quantum Isomer Search}.
\newblock \bibinfo{journal}{\emph{PLoS ONE}} \bibinfo{volume}{15},
  \bibinfo{number}{1} (\bibinfo{year}{2020}), \bibinfo{pages}{e0226787}.
\newblock


\bibitem[\protect\citeauthoryear{Ushijima-Mwesigwa, Negre, and
  Mniszewski}{Ushijima-Mwesigwa et~al\mbox{.}}{2017}]%
        {ushijima2017graph}
\bibfield{author}{\bibinfo{person}{Hayato Ushijima-Mwesigwa},
  \bibinfo{person}{Christian F.~A. Negre}, {and} \bibinfo{person}{Susan~M.
  Mniszewski}.} \bibinfo{year}{2017}\natexlab{}.
\newblock \showarticletitle{Graph partitioning using quantum annealing on the
  D-Wave system}. In \bibinfo{booktitle}{\emph{Proceedings of the Second
  International Workshop on Post Moores Era Supercomputing}}. ACM,
  \bibinfo{pages}{22--29}.
\newblock


\bibitem[\protect\citeauthoryear{Ushijima-Mwesigwa, Negre, Mniszewski, and
  Safro}{Ushijima-Mwesigwa et~al\mbox{.}}{2018}]%
        {mniszewski2018multilevel}
\bibfield{author}{\bibinfo{person}{Hayato~M Ushijima-Mwesigwa},
  \bibinfo{person}{Christian F~A Negre}, \bibinfo{person}{Susan~M Mniszewski},
  {and} \bibinfo{person}{Ilya Safro}.} \bibinfo{year}{2018}\natexlab{}.
\newblock \showarticletitle{Multilevel Quantum Annealing For Graph
  Partitioning}.
\newblock \bibinfo{journal}{\emph{Qubits 2018 D-Wave Users Conference}}
  (\bibinfo{year}{2018}).
\newblock


\bibitem[\protect\citeauthoryear{van Dam, Mosca, and Vazirani}{van Dam
  et~al\mbox{.}}{2002}]%
        {vazirani2002gaps}
\bibfield{author}{\bibinfo{person}{Wim van Dam}, \bibinfo{person}{Michele
  Mosca}, {and} \bibinfo{person}{Umesh Vazirani}.}
  \bibinfo{year}{2002}\natexlab{}.
\newblock \showarticletitle{How Powerful is Adiabatic Quantum Computation?}
\newblock \bibinfo{howpublished}{Proceedings of the 42nd Annual Symposium on
  Foundations of Computer Science, pp. 279-287 (2001)}.
\newblock  (\bibinfo{year}{2002}).
\newblock
\urldef\tempurl%
\url{https://doi.org/10.1109/SFCS.2001.959902}
\showDOI{\tempurl}
\showeprint{arXiv:quant-ph/0206003}


\bibitem[\protect\citeauthoryear{Verdon, Broughton, McClean, Sung, Babbush,
  et~al\mbox{.}}{Verdon et~al\mbox{.}}{2019}]%
        {verdon2019learning}
\bibfield{author}{\bibinfo{person}{Guillaume Verdon}, \bibinfo{person}{Michael
  Broughton}, \bibinfo{person}{Jarrod~R McClean}, \bibinfo{person}{Kevin~J
  Sung}, \bibinfo{person}{Ryan Babbush}, {et~al\mbox{.}}}
  \bibinfo{year}{2019}\natexlab{}.
\newblock \showarticletitle{Learning to learn with quantum neural networks via
  classical neural networks}.
\newblock \bibinfo{journal}{\emph{arXiv preprint arXiv:1907.05415}}
  (\bibinfo{year}{2019}).
\newblock


\bibitem[\protect\citeauthoryear{Von~Luxburg}{Von~Luxburg}{2007}]%
        {von2007tutorial-rs}
\bibfield{author}{\bibinfo{person}{Ulrike Von~Luxburg}.}
  \bibinfo{year}{2007}\natexlab{}.
\newblock \showarticletitle{A tutorial on spectral clustering}.
\newblock \bibinfo{journal}{\emph{Statistics and computing}}
  \bibinfo{volume}{17}, \bibinfo{number}{4} (\bibinfo{year}{2007}),
  \bibinfo{pages}{395--416}.
\newblock


\bibitem[\protect\citeauthoryear{Vo{\ss}, Martello, Osman, and
  Roucairol}{Vo{\ss} et~al\mbox{.}}{2012}]%
        {voss2012meta}
\bibfield{author}{\bibinfo{person}{Stefan Vo{\ss}}, \bibinfo{person}{Silvano
  Martello}, \bibinfo{person}{Ibrahim~H Osman}, {and}
  \bibinfo{person}{Catherine Roucairol}.} \bibinfo{year}{2012}\natexlab{}.
\newblock \bibinfo{booktitle}{\emph{Meta-heuristics: Advances and trends in
  local search paradigms for optimization}}.
\newblock \bibinfo{publisher}{Springer Science \& Business Media}.
\newblock


\bibitem[\protect\citeauthoryear{Walshaw}{Walshaw}{2004}]%
        {Walshaw2004}
\bibfield{author}{\bibinfo{person}{C. Walshaw}.}
  \bibinfo{year}{2004}\natexlab{}.
\newblock \showarticletitle{Multilevel Refinement for Combinatorial
  Optimisation Problems}.
\newblock \bibinfo{journal}{\emph{Annals Oper. Res.}}  \bibinfo{volume}{131}
  (\bibinfo{year}{2004}), \bibinfo{pages}{325--372}.
\newblock


\bibitem[\protect\citeauthoryear{Wang, Hadfield, Jiang, and Rieffel}{Wang
  et~al\mbox{.}}{2018}]%
        {wang2018quantum}
\bibfield{author}{\bibinfo{person}{Zhihui Wang}, \bibinfo{person}{Stuart
  Hadfield}, \bibinfo{person}{Zhang Jiang}, {and} \bibinfo{person}{Eleanor~G
  Rieffel}.} \bibinfo{year}{2018}\natexlab{}.
\newblock \showarticletitle{Quantum approximate optimization algorithm for
  MaxCut: A fermionic view}.
\newblock \bibinfo{journal}{\emph{Physical Review A}} \bibinfo{volume}{97},
  \bibinfo{number}{2} (\bibinfo{year}{2018}), \bibinfo{pages}{022304}.
\newblock


\bibitem[\protect\citeauthoryear{Wu, Di, Dasgupta, Cappello, Finkel,
  et~al\mbox{.}}{Wu et~al\mbox{.}}{2019}]%
        {wu2019full}
\bibfield{author}{\bibinfo{person}{Xin-Chuan Wu}, \bibinfo{person}{Sheng Di},
  \bibinfo{person}{Emma~Maitreyee Dasgupta}, \bibinfo{person}{Franck Cappello},
  \bibinfo{person}{Hal Finkel}, {et~al\mbox{.}}}
  \bibinfo{year}{2019}\natexlab{}.
\newblock \showarticletitle{Full-State Quantum Circuit Simulationby Using Data
  Compression}. In \bibinfo{booktitle}{\emph{Proceedings of the High
  Performance Computing,Networking, Storage and Analysis International
  Conference (SC19)}}. \bibinfo{publisher}{IEEE Computer Society},
  \bibinfo{address}{Denver, CO, USA}.
\newblock


\bibitem[\protect\citeauthoryear{Yung, Casanova, Mezzacapo, McClean, Lamata,
  et~al\mbox{.}}{Yung et~al\mbox{.}}{2014}]%
        {Yung2014}
\bibfield{author}{\bibinfo{person}{M.-H. Yung}, \bibinfo{person}{J. Casanova},
  \bibinfo{person}{A. Mezzacapo}, \bibinfo{person}{J. McClean},
  \bibinfo{person}{L. Lamata}, {et~al\mbox{.}}}
  \bibinfo{year}{2014}\natexlab{}.
\newblock \showarticletitle{From transistor to trapped-ion computers for
  quantum chemistry}.
\newblock \bibinfo{journal}{\emph{Scientific Reports}} \bibinfo{volume}{4},
  \bibinfo{number}{1} (\bibinfo{date}{Jan.} \bibinfo{year}{2014}).
\newblock
\urldef\tempurl%
\url{https://doi.org/10.1038/srep03589}
\showDOI{\tempurl}


\bibitem[\protect\citeauthoryear{Zhou, Wang, Choi, Pichler, and Lukin}{Zhou
  et~al\mbox{.}}{2018}]%
        {zhou2018quantum}
\bibfield{author}{\bibinfo{person}{Leo Zhou}, \bibinfo{person}{Sheng-Tao Wang},
  \bibinfo{person}{Soonwon Choi}, \bibinfo{person}{Hannes Pichler}, {and}
  \bibinfo{person}{Mikhail~D Lukin}.} \bibinfo{year}{2018}\natexlab{}.
\newblock \showarticletitle{Quantum Approximate Optimization Algorithm:
  Performance, Mechanism, and Implementation on Near-Term Devices}.
\newblock \bibinfo{journal}{\emph{arXiv preprint arXiv:1812.01041}}
  (\bibinfo{year}{2018}).
\newblock


\end{thebibliography}
	\appendix
\section{D-Wave Timing Results}
\begin{figure*}
\includegraphics[width=0.5\textwidth]{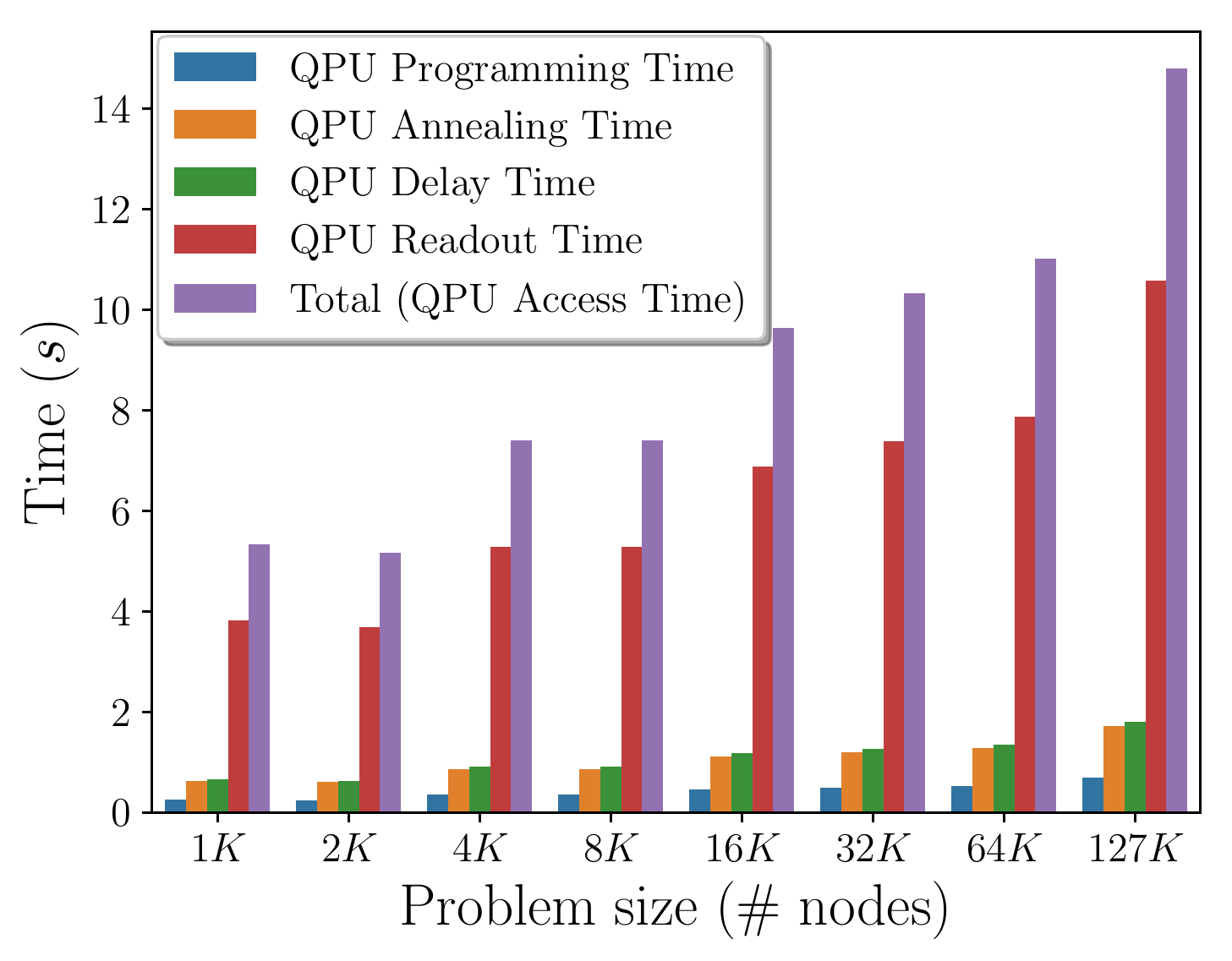}
\caption{D-Wave timing results showing a breakdown of the total time accessing the QPU as the problem size respresented by the number of nodes of the graph increases. }
\label{fig:dwave_timing}
\end{figure*}

For Quantum Annealing on D-Wave, each iteration of the refinement process requires the execution of a single Quantum Machine Instruction (QMI) which includes the QUBO parameters and annealing-cycle parameters sent to the D-Wave system for processing. The total time accessing the QPU (QPU Access time) for a single QMI can be broken down into 4 parts. The first is the one-time initialization step to program the QPU performed at every independent QPU call. Then for each sample requested we have the following times. The annealing time, followed by the time needed to wait for the QPU to regain its initial temperature, referred to as the QPU Delay Time, and lastly the time needed to read the sample from the QPU referred to as the QPU Readout Time. In other words,
\begin{equation}
    qpu\_access\_time=qpu\_programming\_time+N*(annealing\_time+delay\_time+readout\_time)
    \label{qpu_timing}
\end{equation}
where $N$ is the number of samples requested in a single QMI. The timing $N*(annealing\_time+delay\_time+readout\_time)$ is referred to as the QPU Sampling time. In our experiments we request 1000 samples for each QMI and fixed the annealing time to $20 \mu s$ per sample. The annealing time as a user parameter can be either increased or reduced. However significantly reducing this parameter could greatly affect the QPU's ability to find lower energy states, thus affecting the number of iterations of refinement algorithm, and subsequently the entire outcome of the method. In order to test and demonstrate scalability of the proposed approach on the D-Wave system, similar to the experiments presented in Figure~\ref{fig:probsize_scaling}, we varied the number of nodes for a road network graph from approximately $10^3$ nodes up to $10^5$ nodes and computed the maximum modularity of each graph. Similar to the case in Figure~\ref{fig:probsize_scaling} where each QUBO was solved up to optimality we observe a logarithmic scaling in the number of calls to the D-Wave system (number of iterations) as shown in Figure~\ref{fig:probsize_scaling}. Figure~\ref{fig:dwave_timing} gives a breakdown of the total time (Total QPU Access time) for all iterations to compute the modularity for a given input graph where the smallest graph with approximately $1000$ nodes required a QPU Access time of 5 seconds while the largest graph with approximately $ 127,000 $ nodes required a QPU Access time of approximately 15 seconds. However, the majority of the QPU Access time was actually in the QPU Readout time. Requiring less than 2 seconds of total annealing time for 1000 samples is an impressive result for a graph of approximately 127,000 nodes. This time could be significantly trivially reduced by requesting a smaller number of samples per iteration, however this may affect the solution quality. 
Note that the QPU Access time is a linear function in the number of iterations of the refinement method. Due to the heuristic nature of the proposed approach, the total number of iterations by itself is not a strictly increasing function in the number of nodes of the input graph. Thus for example, we observe in Figure~\ref{fig:probsize_scaling}, the number of iterations is not a strictly increasing function and the graphs with $2K$ and $8K$ nodes required a smaller number of iterations than the graphs with $1K$ and $4K$ nodes respectively. This is subsequently observed in the running times as shown in Figure~\ref{fig:dwave_timing}.

\end{document}